\documentclass[11pt]{article}
\catcode`\@=11

\@addtoreset{equation}{section}
\def\theequation{\arabic{section}.\arabic{equation}}
\catcode`\@=11

\def\section{\@startsection{section}{1}{\z@}{3.5ex plus 1ex minus
   .2ex}{2.3ex plus .2ex}{\large\bf}}

\def\eqnarray{\let\@currentlabel=\theequation\refstepcounter{equation}
    \global\@eqnswtrue
    \global\@eqcnt\z@\tabskip\@centering\let\\=\@eqncr
    $$\halign to \displaywidth\bgroup\@eqnsel\hskip\@centering
      $\displaystyle\tabskip\z@{##}$&\global\@eqcnt\@ne
       \hfil${{}##{}}$\hfil
      &\global\@eqcnt\tw@ $\displaystyle\tabskip\z@{##}$\hfil
       \tabskip\@centering&\llap{##}\tabskip\z@\cr}
\def\lefteqn#1{\hbox to 4\arraycolsep{$\displaystyle #1$\hss}}
\def\thesection{\arabic{section}.}

\def\appendix{\setcounter{section}{0}
        \def\thesection{Appendix.}
        \def\theequation{\Alph{section}.\arabic{equation}}}

\long\def\@makefntext#1{\parindent 0cm\noindent
\hbox to 1em{\hss$^{\@thefnmark}$}#1}
\def\IR{{\hbox{{\rm I}\kern-.2em\hbox{\rm R}}}}
\def\IH{{\hbox{{\rm I}\kern-.2em\hbox{\rm H}}}}
\def\IC{{\ \hbox{{\rm I}\kern-.6em\hbox{\bf C}}}}
\def\IZ{{\hbox{{\rm Z}\kern-.4em\hbox{\rm Z}}}}

\newcommand{\beq}{\begin{equation}}
\newcommand{\be}{\begin{equation}}
\newcommand{\eeq}{\end{equation}}
\newcommand{\ee}{\end{equation}}
\newcommand{\bea}{\begin{eqnarray}}
\newcommand{\eea}{\end{eqnarray}}
\newcommand{\bean}{\begin{eqnarray*}}
\newcommand{\eean}{\end{eqnarray*}}
\newcommand{\ba}{\beq\begin{array}{lll} }
\newcommand{\ea}{\end{array}\eeq}
\newcommand{\bi}{\bibitem}

\def\r{\rangle}
\def\l{\langle}

\newcommand{\dis}{\displaystyle}
\def\IC{ {\rm l\hspace{-1.2ex}C} }    %  mine is better....
\def\IZ{{\hbox{{\rm Z}\kern-.4em\hbox{\rm Z}}}}
\def\IR{{\hbox{{\rm I}\kern-.2em\hbox{\rm R}}}}
\begin{document}                           %
                                                                 %
                                                                 %
%%%%%%%%%%%%%%%%%%%%%%%%%%%%%%%%%%%%%%%%%
%     C I T E . S T Y
%     compressed lists of numerical citations: [11-16]
%     see also OVERCITE.STY and DRFTCITE.STY
%
%     Copyright (C) 1989-1992 by Donald Arseneau
%     These macros may be freely transmitted, reproduced, or modified for
%     non-commercial purposes provided that this notice is left intact.
%
%
%  \@citen contains the code that parses the list of names, ignoring
%  spaces after commas, writes the aux file \citation, and formats the
%  number list.  \citen can be used by itself to give citation numbers
%  without the other formatting; e.g., "See also ref.~\citen{junk}."
%
\def\citen#1{%
\edef\@tempa{\@ignspaftercomma,#1, \@end, }% ignore spaces in parameter list
\edef\@tempa{\expandafter\@ignendcommas\@tempa\@end}%
\if@filesw \immediate \write \@auxout {\string \citation {\@tempa}}\fi
\@tempcntb\m@ne \let\@h@ld\relax \let\@citea\@empty
\@for \@citeb:=\@tempa\do {\@cmpresscites}%
\@h@ld}
%
% for ignoring spaces in the input:
\def\@ignspaftercomma#1, {\ifx\@end#1\@empty\else
   #1,\expandafter\@ignspaftercomma\fi}
\def\@ignendcommas,#1,\@end{#1}
%
% For each citation, check if it is defined, if it is a number, and
% if it is a consecutive number    that can be represented like 3-7.
%
\def\@cmpresscites{%
 \expandafter\let \expandafter\@B@citeB \csname b@\@citeb \endcsname
 \ifx\@B@citeB\relax % undefined
    \@h@ld\@citea\@tempcntb\m@ne{\bf ?}%
    \@warning {Citation `\@citeb ' on page \thepage \space undefined}%
 \else%  defined
    \@tempcnta\@tempcntb \advance\@tempcnta\@ne
    \setbox\z@\hbox\bgroup % check if citation is a number:
    \ifnum\z@<0\@B@citeB \relax
       \egroup \@tempcntb\@B@citeB \relax
       \else \egroup \@tempcntb\m@ne \fi
    \ifnum\@tempcnta=\@tempcntb % Number follows previous--hold on to it
       \ifx\@h@ld\relax % first pair of successives
          \edef \@h@ld{\@citea\@B@citeB}%
       \else % compressible list of successives
%         % use \hbox to avoid easy \exhyphenpenalty breaks
          \edef\@h@ld{\hbox{--}\penalty\@highpenalty \@B@citeB}%
       \fi
    \else   %  non-successor--dump what's held and do this one
       \@h@ld \@citea \@B@citeB \let\@h@ld\relax
 \fi\fi%
 \let\@citea\@citepunct
}
%
%%    To put space after the comma, use:
\def\@citepunct{,\penalty\@highpenalty\hskip.13em plus.1em minus.1em}%
%%    For no space after comma, use:
%% \def\@citepunct{,\penalty\@highpenalty}%
%%
%
%  Make \@citex refer to \citen:
%
\def\@citex[#1]#2{\@cite{\citen{#2}}{#1}}%
%
%  Replacement for \@cite.  Give one normal space before the citation,
%  set high penalties for linebreaks,
%
\def\@cite#1#2{\leavevmode\unskip
  \ifnum\lastpenalty=\z@ \penalty\@highpenalty \fi % highpenalty before
  \ [{\multiply\@highpenalty 3 #1% % triple-highpenalties within list
      \if@tempswa,\penalty\@highpenalty\ #2\fi % and before note.
    }]\spacefactor\@m}
\let\nocitecount\relax  % in case \nocitecount was used for drftcite
%%%%%%%%%%%%%%%%%%%%%%%%%%%%%%%%%%%%%%%%%%%%%%%%%%%%%%%%%%%%%%%%%%%%%%%
%%%%%%%%%%%%%%%%%%%%%%%%%%%%%%%%%%%%%%%%%%%%%%%%%%%%%%%%%%%%%%%%%%%%%%%
%                          BODY                                       %
%%%%%%%%%%%%%%%%%%%%%%%%%%%%%%%%%%%%%%%%%%%%%%%%%%%%%%%%%%%%%%%%%%%%%%%
%%%%%%%%%%%%%%%%%%%%%%%%%%%%%%%%%%%%%%%%%%%%%%%%%%%%%%%%%%%%%%%%%%%%%%%%%%%
%%%%%%%%%%%%%%   Giulio's crazy  DEFiNITIONS     %%%%%%%%%%%%%%%%%%%%%%%%%%
%%%%%%%%%%%%%%%%%%%%%%%%%%%%%%%%%%%%%%%%%%%%%%%%%%%%%%%%%%%%%%%%%%%%%%%%%%%
\newcommand{\tPhi}{  {\mbox{\tiny $\!\! \Phi$} } }
%%% 
\newcommand{\pSl}{  \!  \mbox{  \large \em p\raisebox{-.6ex}{{\tiny $\! \lambda$}}}    }
%%%%
\newcommand{\hpSl}{  \mbox{  $\hat{\mbox{ \hspace{-1ex}\large \em p}}$\raisebox{-.6ex}{{\tiny $\! \lambda$}}}   }
%%%%
\newcommand{\dpSl}{  \mbox{  $\dot{\mbox{ \hspace{-1ex}\large \em p}}$\raisebox{-.6ex}{{\tiny $\! \lambda$}}}   }
\newcommand{\etaSpSl}
{  \eta_1\,}
%%%%
\newcommand{\rhoSpSl}
{\rho_1 \,}
%%%%%
\newcommand{\hetaSpSl}
{ \hat{\eta}_1 }
%%%%
\newcommand{\hrhoSpSl}
{ \hat{\rho}_1}
%%%%%
%%%%%%%%%%%
\newcommand{\etaSPhi}
{   \eta_0 \,}
%%%%
\newcommand{\rhoSPhi}
{  \rho_0 \,}
%%%%%
\newcommand{\hetaSPhi}
{ \hat{\eta}_0}
%%%%
\newcommand{\hrhoSPhi}
{ \hat{\rho}_0}

%%%%%%%%
\newcommand{\psiZpSl}
{ \psi^1 \,}
%%%%
\newcommand{\psiZPhi}
{ \psi^0 \,}
%%%%
\newcommand{\psiZPhila}
{\psi^{01} \,}
%%%%%%%%%%%%%%%%%%%%%%%%%%

%%%%%%%%%%%%%%%%%%%%%%%%%%
\newcommand{\LamZpSl}
{\Lambda^1 \,}
%%%%
\newcommand{\LamZPhi}
{\Lambda^0 \,}
%%%%
\newcommand{\LamZPhila}
{\Lambda^{01} \,}
\newtheorem{prop}{Proposition}
%%%%%%%%%%%%%%%%%%%%%%%%%%%%%%%%%%%%%%%%%%%%%%%%%%%
%\input{HD:Giulio:Tex-Projects:header.tex}
\def\itif{\int_{\tau_i}^{\tau_f}}
\def\para{\parallel}
%\newcommand{\pSl}{  \!  \mbox{  \large \em p\raisebox{-.6ex}{{\tiny $\! \lambda$}}}    }
%\newcommand{\dpSl}{  \mbox{  $\dot{\mbox{ \hspace{-1ex}\large \em p}}$\raisebox{-.6ex}{{\tiny $\! \lambda$}}}   }
%%%%%%%%%%%%%%%%%%%%%%%%%%%%%%%%%%%%%%%%%%%%%%%%%%%%%%%%%%%%%%%%%%%%%%%%%%
%%%%%%%%%%%%%%%%%%%%%%%%%%%%%%%%%%%%%%%%%%%%%%%%%%%%%%%%%%%%%%%%%%%%%%%%%

\begin{titlepage}
\vspace{.5in}
%\begin{flushright}
%Los Alamos  \\
%IEEC-Jun-97\\
%gr-qc/???????\\
%June 1997\\ {\tiny (This version: \today) }
%\end{flushright}
\vspace{.5in}
\begin{center}

{\Large\bf Four approaches to quantization of the relativistic particle}\\
\vspace{.4in} 
%{Emil ~M{ottola}\footnote{\it email: emil@pion.lanl.gov}\\
%        {\small\it Theoretical Division Group, T-8, MS-B285 }\\
%       {\small\it Los Alamos National Laboratory}\\
%        {\small\it Los Alamos, NM 87545, USA}}\\
% \vspace{1ex}
   {Giulio ~R{uffini}\footnote{\it email: ruffini@ieec.fcr.es}\\
        {\small\it Earth \& Space Sciences Group}\\
        {\small\it Institut D'Estudis Espacials de Catalunya, CSIC Reseach Unit}\\
{\small\it Edif. Nexus, 204} \\
{ \small\it Gran Capit\`a,  2-4 \\ }
        {\small\it 08034 Barcelona, Spain}
        } 
 \end{center}

 \vspace{.2in}
\begin{center}
{\large\bf Abstract}
\end{center}
\begin{center}
\begin{minipage}{5.4in}
{\small    The connection between four different approaches to quantization of the relativistic particle is studied: reduced phase space quantization, Dirac quantization, BRST quantization, and (BRST)-Fock quantization are each carried out. The connection to the BFV path integral in phase space is provided. In particular, it is concluded that  that the full range of the lapse should be used in such path integrals. The relationship between all these approaches is established.}
    \end{minipage}
    \end{center} 
\end{titlepage}

\section{Introduction}
\setcounter{footnote}{0}
Investigating  the problems that arise in the quantization of the parametrized relativistic particle 
may not seem like a very promising task.  After all,  
it has been known for a long time that to implement relativistic 
covariance a quantum field theory for this system needs to be developed, 
so we expect any ``first-quantization'' approach to ultimately fail.  
Although this is a reasonable consideration, it is also 
true that the  parametrized relativistic particle, 
 a constrained system, bears a close resemblance 
 to more complex and interesting systems, such as string theory, 
mini-superspace, and gravity. All of these  are
constrained systems, and many of  the deep issues that 
arise trying to quantize them play a role in the quantization of
 the parametrized relativistic particle \cite{Hal2}. 
 In addition, it is
not obvious that the reasons that have led us to abandon a ``first-quantized''
theory for the relativistic particle and turn to quantum field theory,
 will be valid in the case of gravity. For these reasons, every effort
 should be made to completely understand what the different methods for
 quantization of a constrained system yield in the case of the 
 parametrized relativistic particle.

One of the  problems that these systems share is that of the
 range of the ``lapse'', which is especially apparent  in the covariant path integral 
approach to quantum gravity \cite{Mottola}.  The problem there is that a coordinate space covariant path
integral for gravity can be constructed using string theoretic methods, but one of the integration
variables, called the lapse, has an undefined
range.   Relativistic covariance dictates only  that the range should be either $(0,\infty)$ or
$(-\infty,\infty)$.   If it were possible to derive this
path integral from a Hilbert space formalism, the range of integration of 
all the variables would be completely fixed.  Understanding 
the Hilbert space structure  of a quantum theory is essential 
to make sense of an object like  a path integral.

The general 
action for the relativistic particle in a curved spacetime and in a 
background electro-magnetic field is given by
\beq 
A =  \itif  d\tau  \left(-m\sqrt{g(x^\alpha ) _{\mu  \nu}   
{dx^\mu 
\over d
\tau}  {dx^\nu \over d\tau}} - e {dx^\mu \over d\tau}A_\mu 
(x^\alpha )  \right) ,
\label{eq:action1}
\eeq
or by  the alternate form
\beq \label{eq:action2}
A' =  \itif  d\tau  \left({1\over \lambda 
(\tau )}   g(x^\alpha ) _{\mu  \nu}{dx^\mu \over d\tau } 
  {dx^\nu \over d\tau }  + m\lambda (\tau ) - e 
{dx^\mu \over d\tau }A_\mu (x^\alpha )  \right) .
\eeq
The boundary condition needed for an extremum are $x^\mu (\tau_i)=x^\mu_i$ and $x^\mu (\tau_f)=x^\mu_f$.
Let us begin studying   equation \ref{eq:action1}---in either case the end product
is the same, as we will show in a moment. This action is 
invariant under reparametrizations that do not affect the
boundaries:  $\tau  \longrightarrow  f( \tau ) $ with $f( 
\tau_i)=
\tau_i$ and $f( \tau_f)=  \tau_f$,  and with $ df / d \tau  >0$. If $e=0$, 
the action is also invariant under reparametrizations satisfying  $\tau  \longrightarrow  f( \tau ) $,
 with $f( \tau_i)=\tau_f$ and $f( \tau_f)=  \tau_i$,  and with $ df / d \tau  < 0$.
We can think of the full reparametrization group as  the direct product 
of $Z_2$ and the group of reparametrizations connected with the identity, and of this action as carrying
 an unfaithful representation of
the $Z_2$ part of the reparametrization group. The action of the  $Z_2$
part of the reparametrization group can be described by two types of 
reparametrization functions: $f_+$, which maps $\tau_i$ and $\tau_f$
into
themselves, and $f_-$, which maps $\tau_i$ into $\tau_f$ and vice versa.
The group multiplication is rules are
\beq
Z_2
= \left\{  \begin{array}{ll}
f_+ \cdot f_+ &= f_+\\
f_+\cdot f_- &= f_-\\
f_- \cdot f_- & = f_+ 
\end{array}
\right.
\eeq
The full reparametrization group is given by ${\cal G} = Z_2 \otimes {\cal F}_+$
where $ {\cal F}_+$ denotes the part connected to the identity. The action 
in equation \ref{eq:action2} carries a faithful representation of ${\cal G}$, with
$\lambda(\tau) \rightarrow \dot{f} \lambda(f)$, for an  arbitrary $e$. 

 Defining the momenta in the usual way we 
find---because of the reparametrization invariance---the first class constraint
\beq\label{eq:constraint}
\Phi = (p_\mu -A_\mu) g^{\mu \nu} (p_\nu -A_\nu) -m^2 
\approx 0 , 
\eeq
and  a zero hamiltonian, $ H \equiv 0$. 
The
same equations of motion can be obtained from the so-called first 
order
action in the phase space coordinates
\beq
F=\itif d\tau(p_t \dot{t} + p_\mu \dot{x}^\mu - v\Phi), 
\eeq
which is invariant under the gauge transformations 
 $
\delta x^\mu = \epsilon (\tau) \{x^\mu, \Phi\}$, $ \delta p_\mu = 
\epsilon(\tau)\{p_\mu,\Phi\}$, and 
$\delta v = \dot{\epsilon}(\tau)$,
as long as the gauge parameter vanishes at the boundaries, i.e.,  
$\epsilon 
(\tau_i) = \epsilon (\tau_f) =0$. It is not hard to see also
that this symmetry is the same as the  one in the 
Lagrangian form, with the identification $f(\tau) = \tau +\epsilon 
(\tau)$. 
The  
restriction in the gauge freedom at the boundaries is not usually present 
in the gauge theories. 
As explained in reference \cite{Teitelpap}, this twist in the 
concept of 
invariance is a 
consequence of the form of the constraint, which is non-linear in the 
coordinates conjugate to what is fixed at the boundaries, i.e., the 
momenta. 
Under the above gauge transformation, the first order action changes, 
as a 
boundary term appears:
\beq
A \longrightarrow A + \epsilon (\tau)\left.  (p_i {\partial \Phi \over 
\partial 
p_i} - 
\Phi) 
\right| _{\tau_i}^{\tau_f}.
\eeq
This term vanishes if the constraint $\Phi$ is linear in the 
momenta.  We will discuss this point further below.

Using the action in equation \ref{eq:action2}, the 
situation is just a bit
more complicated. First 
we find the constraint $\pSl \approx 0 $, since the action is 
independent of $\dot{\lambda}$. The hamiltonian is not zero, $ 
H' \equiv \lambda \Phi$. The condition 
$\dpSl =0$, which is needed to insure that the constraint
 is preserved by the dynamics  \cite{Dirac},
implies 
$\Phi \approx 0$, i.e., the constraint in equation \ref{eq:constraint}
 appears here as a 
secondary 
constraint. The extended hamiltonian is  given by
$
H_E = \lambda\,  \Phi + u \, \Phi + w\, \pSl .
$
The dynamics that do not involve the trivial degree of freedom $\lambda$ are thus
described as in the earlier case.

The important features  of the parametrized relativistic particle system are:
\begin{itemize}
\item It is a constrained system with a first class constraint. 

\item It is a parametrized system. % This tends to confuse issues, but is not really a serious problem.  
       The actions for such systems are not entirely ``gauge'', since they do not have gauge freedom at the 
             boundaries.

\item The constraint surface has a non-trivial topology---it is disconnected and non-compact---and
the constraint cannot be made into a momentum by a canonical transformation.  
%This is really the source of all problems. 

\end{itemize}
When looking at this list it is useful to keep in mind the parametrized non-relativistic case.
It is indeed possible to construct a parametrized theory of the non-relativistic particle, 
by making $t$ into a dynamical variable \cite{Dirac,TeitelBook}. The action is given by
\beq
\label{eq:unc}
S = \itif d\tau \; L = \itif  d\tau \; \left( { m \over 2}{\dot{x}^2\over \dot{t} } -\dot{t}V(t,x) \right)  
\eeq
%The solution to the equations of motion is given by $t=f(\tau)$, $x=v f(\tau) + x_0$, 
%with $f(\tau_{i,f})=t_{i,f}$, $x(\tau_{i,f})= x_{i,f}$.
  The constraint in this 
case is given by $p_t+R(t,x,p_x)\approx 0$, with $R(t,x,p_x)=p_x^2/(2m)+V(t,x)$.
 This system is not hard to quantize, although
it shares the  features in the first two items of the 
 list with the relativistic case.  The reason is  that the constraint
has a trivial topology and can be made into a momentum by a canonical transformation. 
It is well-known that all approaches to quantization of a constrained system 
are equivalent when it is possible to make the constraints into 
momenta by a canonical transformation.  The last item in the list is the one 
which really makes things difficult for the relativistic particle, as we will see.

In all these aspects the relativistic particle is similar to string theory, 
mini-superspace and gravity, and we  should understand 
this simple  system %, such as  the range of the lapse, 
 before attempting to understand  more complicated 
ones.  For example, we will  shown that when the  electric field is divergence-less,
 unitary single-particle covariant quantization of the relativistic particle
is possible. %, as we expect from  quantum field theory \cite{itz}. 
The result is
 not very surprising, but  
relevant may be   relevant to  
quantum  gravity.  
What are  the analog conditions to ``covariance'' and  an electric field with no divergence, if any, 
in quantum gravity?

Aside from this, 
the parametrized relativistic particle is 
the simplest non-trivial example in which to study the equivalence of the different 
quantization methods available today: reduced phase space quantization, 
Dirac quantization, BRST quantization, and (BRST)-Fock quantization.  In fact, these methods are only
well-defined for the simplest cases, and studying this system may indicate 
 the proper way to their generalization.  

Let us now look more closely at some of the more complicated systems with 
features  in common with the  parametrized relativistic particle.
As mentioned, mini-superspace
models are mathematically very similar to the relativistic
particle in a curved background.  As an example, consider the 
Robertson-Walker model 
described by the metric
$
ds^2 = {- N^2(\tau) \over q(\tau) } dt^2 + q(\tau) d\Omega_3^2
$, 
where $d\Omega_3^2$ is the metric on the unit three-sphere 
\cite{Hal4}.
The  constraint that appears in this theory is 
\cite{Louko87}
$ 4p^2 -\Lambda q +1 \approx 0$. 
Although similar to the constraint in  \ref{eq:constraint}, this constraint is actually more closely
related to  the one that appears in the parametrized non-relativistic particle: notice that
the constraint surface is simply connected, and from this point of view quantization should
not be very difficult. 
More general  mini-superspace models is 
\beq
S_{M} = {1\over 2} \itif d\tau \,    \left(    {g_{AB} \dot{Q}^A \dot{Q}^B 
\over 
N}   + N U(Q)   \right), 
\eeq
where the signature of $g_{AB}$ is Lorentzian. This action is  equivalent to
$
S'_{M} = - {1\over 2} \itif d\tau\, \sqrt{ U(Q) g_{AB} \dot{Q}^A \dot{Q}^B }
$. 
 In this homogeneous 
cosmological models  the lapse $N$  is essentially the 
time-time 
component of the metric, and the  parameter $Q^0$ characterizes
the scale of the universe. The other parameters describe spatial 
anisotropies.  The associated constraint in this model is  
\beq
\Phi_M \equiv P_A P_B g^{AB} - U(Q) \approx 0, 
 \eeq
which looks like the one in equation \ref{eq:constraint}  with  
with a 
coordinate dependent mass.

The action for (3+1)-dimensional gravity is given by
\beq
S_H = \int d^4x \, \sqrt{{}^{\scriptscriptstyle(4)}\!g} \, \left(
{}^{\scriptscriptstyle(4)}\!  R-2\Lambda \right).
\eeq
 In the Arnowitt-Deser-Misner
formalism \cite{ADM} it is assumed that the topology of space-time is of the form
$\IR\!\times\!\Sigma$, and with the metric expressed by $
ds^2 = N^2 dt^2 - g_{ij}(dx^i + N^i dt)(dx^j + N^j dt)
$ 
the action  becomes, in the canonical formalism ($g$ and $R$ stand for the 3-geometry  metric and curvature in the time
slices, unless specified otherwise by a $(4)$ superscript)
\beq
S_H = \int dt\int_\Sigma  d^3x \, ( \pi^{ij} \dot{g}_{ij} - N^i {\cal H}_i - N {\cal
H}) , 
\eeq
where the constraints are  ${\cal H}_i, {\cal H}\approx 0$, with
\beq
{\cal H}_i = -2 \nabla_j \pi^j_i , \:\:\:\:\: 
{\cal H} = G_{ijkl}\, \pi^{ij} \pi^{kl} - \sqrt{g}(R-2\Lambda), 
\eeq
and 
\beq
 G_{ijkl}= {1\over 2\sqrt{g}} (g_{ik} g_{jl}+  g_{il} g_{jk}  -g_{ij} g_{kl}  ), 
\eeq
and where $\pi^{ij}(x)$ is the momentum conjugate to $g_{ij}(x)$,
$\{g_{ij}(x)  ,\pi^{kl}(x')\} = \delta_i^k \,\delta_j^l\, \delta^3(x-x')$.
The linear constraints generate the space diffeomorphisms, while the 
constraint ${\cal H}\approx 0$  generates the dynamics. The variable  $N$ is called the 
lapse function, and is akin to $\lambda$ in the relativistic particle case.
The linear constraints   can 
be treated in the reduced phase space context, but  the appearance of the  quadratic
constraint is much more difficult to treat. It is also possible in this case to
 write a ``square-root'' action 
which is very reminiscent of equation   \ref{eq:action1}
\cite{Hartle86},
\beq
S[g_{ab},N^a] = \int d\tau \int_{\Sigma} d^3 x \sqrt{ gR(\dot{U}_{ab}\dot{U}^{ab} 
- U^2)}, 
\eeq where $U_{ab} \equiv \dot{g}_{ab} - N_{a;b} - N_{b;a}$.
A simplified version of this theory is   (2+1)-dimensional gravity. This
is a remarkable system, because 
it is described by a finite number of degrees of freedom---the
constraints eliminate almost all of them.
As discussed in \cite{Carlip93,Carlip941}, in this case
the different quantization schemes available seem to yield
substantially different theories. For example \cite{Moncrief,Carlip90},
it is possible, through a proper choice of ``extrinsic time'', to reduce
explicitly  the theory classically and  quantize the finite number of remaining
degrees of freedom. The Chern-Simons approach to (2+1)-gravity is
also of the ``constraint then quantize'' kind, and  
the equivalence of these methods has been  established, at least for simple topologies \cite{Carlip90}. 
 On the other hand,  the Wheeler-DeWitt approach \cite{Carlip941}, i.e., Dirac quantization,
is  much less understood.

 These examples are meant to 
highlight the similarities of these systems and  the parametrized relativistic particle, and
convince the reader that quantization methods  should first  be
 tested and completely understood 
 in such a simple theory. Some of the problems that may arise 
  are likely to be already present in this toy model, 
making  their solution  more
 readily apparent.

We will begin by showing that reduced phase space quantization is not possible in general 
backgrounds, since
it leads to the the square-root 
Schr\"{o}dinger equation. The behavior of the square-root Schr\"{o}dinger equation
under the Lorentz group has already been studied by Sucher \cite{Sucher}, 
who showed that in general it is not covariant.   We will show, however, 
 that this equation
 transforms properly under Lorentz boosts when the electric field is divergence-less.  In the next
section we will discuss full Dirac quantization of the relativistic particle, and 
define an inner product following some simple considerations.  Then,  we will discuss 
 BRST quantization and define the BRST inner product. Finally,  we will generalize and use
(BRST)-Fock quantization. 

%%%%%%%%%%%%%% 1
\section{Reduced phase space quantization}
Reduced phase space quantization, a
``constrain, then quantize'' approach, provides a clear conceptual
framework for quantization of those   systems in which it is possible to make 
the constraints into momenta by a canonical transformation.  
Notice that if is is possible to perform such a canonical transformation
then there exists  a 
well-defined function in phase space that satisfies $\{\xi, \Phi\}=1$.  In the case of the
relativistic particle it is impossible to find such a function.  The reader can check that 
in the free case $\{\xi, \Phi\}=0$ at $p_\mu=0$, for any non-singular $\xi$. 
As a consequence,  in the  reduction process 
we will encounter problems---we will end up with two classical theories instead of one. This,
of course, is due to  the disconnectedness of the constraint surface. For simplicity we 
analyze first the 2-dimensional free case.  

To ``reduce'' the phase space we add another constraint to the
system, $\Upsilon\approx 0$, the so-called gauge-fixing function, and work with the hamiltonian
$H_{E'}=v(\tau) \Phi+ w(\tau)\Upsilon$. We will  use here a  $\tau$-dependent 
gauge, 
$ \Upsilon_f = t - f(\tau) $ (this type of gauge-fixing function is 
sometimes called ``canonical'', because it does not  involve  lagrange multipliers).
Next, we  need to insure that
the constraint and the gauge-fixing term  are preserved under the dynamics. 
The condition $\Phi\approx 0$  implies $w=0$, while   
$
\dot{\Upsilon}_f = {\partial \Upsilon_f  / \partial \tau} +
 \{\Upsilon_f, H_{E} \} \approx 0
$  implies $ v=\dot{f}/2p_t$, so 
\beq
H_{E'} = { \dot{f}\over 2p_t} \Phi  =  {\dot{f} \over 2p_t} ( p_t^2-  p_x^2-m^2).
\eeq
The equations of motion in this gauge are
$  \dot{t} = \dot{f}$, $  \dot{x}  = -{\dot{f} p_x / p_t} $, 
$ \dot{p}_t   = 0$ and $\dot{p}_x = 0$. 
We can rewrite the constraint as $
p_t =  \pm\sqrt{ p_x^2 + m^2}$ , and also rewrite
\beq
\label{eq:non-rel-red}
\dot{x}  = \pm\{x, \dot{f}\sqrt{ p_x^2 + m^2}\} =\pm\dot{f}\{x,\sqrt{ p_x^2 + m^2}\}. 
\eeq
The system is effectively reduced to the coordinates $x, p_x$, with a 
hamiltonian 
$
H_f =  \pm \dot{f}\sqrt{ p_x^2 + m^2}
$. 
 Notice that the hamiltonian comes with two signs: thus propagation
back and forth in time are both  described in this system, depending on the initial conditions of
the particle.  If the initial conditions are such that the particle is in the positive branch of the constraint
it will stay
there, and evolve with the positive root hamiltonian.  Else, it will move back in time---it will
be an anti-particle.

Equivalently, we can use a Dirac bracket formulation to analyze the reduced phase space.  
The bracket  computation with  this gauge is easily done, since $\{ 
\Upsilon_f, \Phi\} = 2p_t $: $  \{A,B\}_* =  $
\beq
\{A,B\} - ( \{ A,\Upsilon_f \},\{A, \Phi \}) 
\left( \begin{array}{cc} 0 & -{ \{ 
\Upsilon_f, \Phi\}^{-1}}  \\ { \{ 
\Upsilon_f, \Phi\}^{-1}} & 0 \end{array}   \right)
\left( \begin{array}{c} \{\Upsilon_f,B\} \\ \{\Phi ,B\} \end{array} 
\right)           ,              
\eeq
and this yields
\beq
\{x,p_x\}_* = 1,\: \{t,p_t\}_* = 0,\:
\{x,p_t\}_* = \pm \{ x,  \sqrt{p_x^2 + m^2}\}, 
\eeq
with all others zero.  The appearance of the two signs in the bracket structure is a clear sign of trouble.
The dynamics are given by \cite{mythesis}
\beq
\dot{F}= {\partial F \over \partial \tau} + {\partial \Upsilon \over \partial \tau}
{\{ F,\Phi \} \over \{\Upsilon, \Phi\} }.
\eeq
It is important to note that equation \ref{eq:non-rel-red}
is gauge-dependent.  This is because  we have not selected a set of observables, i.e., 
phase space functions that have zero Poisson
bracket with the constraint. Had we done so we would have found that their $\tau$-derivative is zero,
 since as observables they would also commute with the reduced phase space hamiltonian, $H_{E'}$.
It helps to recall here that the action we start with, in equation \ref{eq:action1}, 
 is not fully  gauge-invariant, so it
is not surprising that if we now insist on total gauge-invariance, we do not get back to 
the point that we started with.  This is due to the way in which we have defined the problem----with
gauge-dependent boundary conditions.

In the  free parametrized non-relativistic case this point is especially
transparent. Using a canonical transformation 
we can explicitly separate  the set of physical coordinates, call them $q,p$, 
from the gauge coordinates, $Q, P$ (for notational simplicity we discuss the 2D case),
\bea \label{eq:observables}
Q= t-t_\lambda,    \;\; P= p_t + {p_x^2/ 2m} &\:\:&  \mbox{ ({\em gauge degrees 
of 
freedom})} \nonumber \\ 
q= p_x (t-t_\lambda)-mx ,  \;\;  p=- p_x/m   & \:\:& \mbox{ ({\em physical degrees 
of freedom})}
\eea
After this canonical transformation, one might 
expect that the original action may  be rewritten as $S=\int_{\tau_i}^{\tau_f} d\tau\, 
(\dot{q}p +\dot{Q}P -\omega\Phi) $, but
this is incorrect: a gauge-dependent surface term (the generator of the canonical transformation) 
is missing. The boundary conditions used in the original action,  $x^\mu(\tau_{i,f})=x^\mu_{i,f}$,  
do not involve the physical coordinates
of the theory only:  $Q$ and $-q/m-pQ$ are fixed at the boundaries. The 
above  action does not  have an extremum with such boundary conditions; a 
surface 
term needs to be added. Indeed, 
consider the variation of the action:
\beq
\delta S = \itif( (\dot{Q}- w) \delta P - \dot{P} \delta Q -  P \delta w +
\dot{q}\delta p  -  \dot{p}\delta q ) d\tau     
+\left. (P\delta Q+p\delta q)  \right|_{\tau_i}^{\tau_f}
\eeq
With  boundary conditions on $x,t$, the surviving surface term is 
\beq 
\left. (p\delta q)\right|_{\tau_i}^{\tau_f}
 = - \left. p_x (t-t_\lambda) {\delta p_x\over m}\right|_{\tau_i}^{\tau_f}\eeq
To eliminate it we need to add a surface term   to 
the 
action,  $ \left.  B
\right|_{\tau_i}^{\tau_f} 
$, with 
$ \left. (p\delta q + \delta B) \right|_{\tau_i}^{\tau_f} =0 $. 
A solution is  $B = (t-t_\lambda) {p_x^2\over 2m} = {m \over 2}Q p^2$.
 This term  is the origin of the  lack of gauge invariance at the 
boundaries  comes from, as it depends on $Q$. Let us summarize: there is 
nothing special about the action---or 
the  constraint, as it is simple  enough, $P\approx 0$.  However, 
our insistence on peculiar boundary  conditions 
makes the addition of a gauge dependent boundary term necessary 
for  the  existence of an extremum
(the generator of  the time-dependent canonical transformation needed  to go to
constant-of-the-motion coordinates,  the Principal Function), and  gauge invariance at the 
boundaries is  lost.
   This is not a serious problem, it is just a language problem. 
If we agree to use boundary condition using constant-of-the-motion coordinates we can leave the
surface term out and work with a fully gauge invariant theory (with identical results).
This is the action in  reference~\cite{Teitelpap}.

The proper way to thing about $t$ is as the
coordinate of a reference object we call clock.  Of course, this coordinate is pure gauge,
an arbitrary function of $\tau$. Nonetheless,  it is 
not obligatory  to make it disappear from the formalism.
For instance, we are allowed to say ``when the clock is at $t=t_\lambda$, $x=x_\lambda$'',
or to rewrite  equation 
\ref{eq:non-rel-red} in the form $dx/dt=\pm \sqrt{p_x^2+m^2}$.  
The rigorous way to implement this is to work only with observables, such as the ones in 
equation \ref{eq:observables}.   We can  recover the $t$ coordinate  in
the  un-parametrized formalism not through $\tau$ or $t$, but through $t_\lambda$.  To summarize: properly
speaking, we should only use observables, and recover the concept of time through a shifting
between these observables (using the $p_t$ generator).  
Shifting  between observables can be done by choosing different
gauges if we do not restrict the theory to observables, which is what we have done here.  
Although this is uglier, it is how we started with in the action.

To quantize the reduced phase space 
we must now choose a sign for the hamiltonian (this choice is equivalent to a
choice of  a ``branch'' of the constraint at the beginning of the reduction process). The 
system is reduced to the spatial coordinates
and momenta, and  a Schr\"{o}dinger equation with  a square-root hamiltonian 
(e.g., choosing the negative  root, and with   the gauge $t=\tau$),
\beq  \label{Schr}
i \partial_0 \Psi({x}^{\mu})  = {H}(A^\nu ({x}^{\mu}) , \partial_\mu)) \Psi({x}^\mu) ,
\eeq
with ${H}= \sqrt{D^iD_i +m^2} - A_0$, and $ D_\mu = \partial_\mu - iA_\mu$.
This equation also appears in Dirac quantization (as the condition for physical states)
if a branch of the constraint is 
chosen. It is not hard to see
that the choice of different gauges is equivalent to performing a ($\tau$-dependent) canonical 
transformation in the reduced
phase space, or unitary transformations in the quantum theory 
(leading to different ``pictures'' of quantum mechanics) \cite{mythesis}. 
Let us  see, however,  that 
in the process of decomposing the constraint we have lost a symmetry that was 
present 
in the original action: covariance.  
How should this equation  behave under Lorentz
transformations? The wave-function must transform as
     a relative scalar 3-density of
weight 1/2, if we
are to construct  a probability  interpretation using the usual inner product.  
Thus, under a
change of coordinates  the 
wave-function changes by $\Psi(x^\mu)\longrightarrow \gamma^{1/2}  \Psi(\Lambda^{-
1\mu}_\nu
 x^\nu)$, and we must check that under this 
change the wave-function still satisfies the Schr\"{o}dinger equation. 
%If there is a background field,
% the correct
%covariance statement for an equation of the form 
%$D({\partial\over \partial x^\mu}, A_\mu 
%(x^\mu))
%f(x^\mu)=0$ is that this equation imply that  
%\beq
% D({\partial\over \partial x^\mu}, 
%\Lambda A_\mu
%(\Lambda^{-1}  x^\mu)) \: f(\Lambda^{-1}  x^\mu)=0 .
%\eeq
%In other words, relabeling the  variables, $\Lambda^{-1} x \rightarrow x$,
% we need to have
%\beq
%D(\Lambda_\nu^\mu{\partial\over \partial x^\nu}, 
%\Lambda_\nu^\mu A_\nu
%(x^\mu)) \: f(x^\mu)=0 .
%\eeq
Given   equation \ref{Schr}, we need 
\beq \label{boost}
i \Lambda_0^\mu \, {\partial \over \partial x^{\mu}}
 \Psi({x}^{\mu})  = {H}(\Lambda^\mu_\nu A^\nu ({x}^\nu),
\Lambda^\mu_\nu \partial_\mu) \Psi({x}^\mu) .
\eeq
%Before checking  the validity of equation \ref{boost} consider for
% simplicity the scalar  equation ($i=1,2,3$), 
%$
%|p_0| = \sqrt{-p^i p_i  + m^2}
%$. 
%This equation is equivalent to $p_0^2 - p_i^2 = m^2$ of course, so if
%it is true in one frame of reference it will hold in all of them. Let
%us check this explicitly by boosting both sides,
%$
%|\Lambda^\mu_0 p_\mu | = \sqrt{ -\Lambda_i^\mu p_\mu \Lambda^i_\nu p^\nu 
%+m^2} 
%$ 
%\beq =\sqrt{-\Lambda_\alpha^\mu p_\mu \Lambda^\alpha_\nu p^\nu 
%+\Lambda_0^\mu p_\mu \Lambda^0_\nu p^\nu +m^2} 
%= \sqrt{\Lambda_0^\mu p_\mu \Lambda^0 _\nu p^\nu} ,
%\eeq
%so it holds after a boost, as it should, and that 
For simplicity, suppose first  that
 $\Psi(x^\mu)$  satisfies the free case of  equation~\ref{Schr} ($A_\mu=0$). 
 We  have
to check that 
\beq
i \Lambda_0^\mu \, {\partial \over \partial x^{\mu}}
 \Psi({x}^{\mu}) 
= \sqrt{ \Lambda_i^\mu \partial_\mu \Lambda^i_\nu \partial^{\nu} +m^2}\, 
\Psi({x}^\mu) .
\eeq
Rewriting the right hand side, we obtain
\bea
\sqrt{ \Lambda_i^\mu \partial_\mu \Lambda^i_\nu \partial^{\nu} +m^2}\, 
\Psi({x}^\mu) &=&
\sqrt{ \Lambda_\alpha^\mu \partial_\mu \Lambda^\alpha_\nu \partial^{\nu} 
-\Lambda_0^\mu \partial_\mu \Lambda^0_\nu \partial^{\nu} +m^2}\, 
\Psi({x}^\mu) \nonumber \\
&=&
\sqrt{-\Lambda_0^\mu \partial_\mu \Lambda^0 _\nu \partial^{\nu}}\, 
\Psi({x}^\mu) ,
\eea
which is consistent. Notice that the crucial element in the derivation  was that the square-root 
equation imply the Klein-Gordon equation,
\beq
i \partial_0 \Psi(x^\mu) = \sqrt{\partial^i \partial_i  + m^2}\, \Psi(x^\mu) 
\Longrightarrow
[\partial_\mu \partial^\mu + m^2]\, \Psi(x^\mu) =0 \eeq
which is true because\beq
\partial_\mu \partial^\mu + m^2 = \left(\partial_0 - \sqrt{\partial^i 
\partial_i +m^2}\right)\left(\partial_0 + \sqrt{\partial^i \partial_i+m^2}\right),
\eeq
since  $
[\partial_0, \partial_i] = 0$. % We also needed 
%$ [\partial_\mu, \Lambda_0^\nu \partial_\nu ]=0 $.

%\subsubsection{Electromagnetic interaction}
\setcounter{footnote}{0}
Interaction with a background electro-magnetic field  appears through
the definition of covariant derivatives (minimal coupling), $D_\mu = \partial_\mu - iA_\mu$, 
\beq \label{eq:el}
i D_0 \Psi(x^\mu) = \sqrt{D ^i D_i + m^2}\, \Psi(x^\mu) .
\eeq
 One can show \cite{Samarov} that this equation is $U(1)$
gauge covariant, meaning that if $\psi(x^\mu)$ is a solution and one makes the changes 
\beq
\psi(x^\mu)  \longrightarrow   e^{-ie\Lambda(x^\mu) }\; \psi(x^\mu), \:\:\:\: 
A_\mu\longrightarrow A_\mu +\partial_\mu \Lambda,
 \eeq
the equation is still valid.  In reference 
\cite{Sucher} it was shown that there exist electro-magnetic fields for 
which the square-root equation does not transform properly. Let us 
see what requirements the electro-magnetic must  satisfy in order to
preserve covariance.
As before, we have to check that
\bea
i \Lambda_0^\mu \, {D_0 }
 \Psi({x}^{\mu}) 
&=& \sqrt{ \Lambda_i^\mu D_\mu \Lambda^i_\nu D^{\nu} +m^2}\; \Psi({x}^\mu)  \nonumber\\
&=& \sqrt{ \Lambda_\alpha^\mu D_\mu \Lambda^\alpha_\nu D^{\nu} 
-\Lambda_0^\mu D_\mu \Lambda^0_\nu D^{\nu} +m^2}\; \Psi({x}^\mu) \nonumber \\
&=& \sqrt{D^{2} +m^2-\Lambda_0^\mu D_\mu \Lambda^0 _\nu D^{\nu}}\;  \Psi({x}^\mu). 
\eea  
A sufficient condition for consistency 
is that   the electric field be divergence-less,% vanish in some frame,   
\beq \label{eq:condition}
[D_0,D_iD^i]= \nabla\cdot\vec{E}=0 ,%[D_0, D_i]= F_{0i} = E_i ,
\eeq
 as we  will now show.  Notice that this is a covariant statement: if it holds
in one frame, it  holds in all of them, because what all it says  is that there are
no charges (but one can check this explicitly).  
If this condition is met  the Klein-Gordon operator  decouples,
\beq
[\left(D_0 \pm \sqrt{D^i D _i +m^2}\right)\left(D_0 \mp \sqrt{D ^i D_i+m^2}\right)]_x \, \phi(x) =
\left[D^2 +m^2\right]_x\, \phi(x) .
\eeq
Hence,  
if $\phi(x)$ is a solution of equation \ref{eq:el}, it also satisfies the
the Klein-Gordon equation, $(D^2 +m^2)\phi=0$.
%\beq
%  [\left(D_0 \pm \sqrt{D^i 
%D _i +m^2}\right)\left(D_0 \mp \sqrt{D ^i D_i+m^2}\right)]_x \; \phi(x) =0.
%\eeq
% Moreover, because this equation is relativistic we also know that
% \beq
% [\left(D^\Lambda_0 \pm \sqrt{D^{\Lambda i} 
%D^\Lambda _i +m^2}\right)\left(D^\Lambda_0 \mp \sqrt{D^{\Lambda i} D_i+m^2}\right)]_x \; \phi(x) =0,
%\eeq
%where $D^\Lambda_\mu= \Lambda^\nu_\mu D_\nu$. Writing out this equation for the 
%two signs and subtracting yields
%\bea
%[ D^\Lambda_0 , \sqrt{D^{\Lambda i} D^\Lambda_i+m^2}]_x \; \phi(x)  &=&
% [ D^\Lambda_0 , m^2 ( 1+  {1\over 2} {D^{\Lambda i} D^\Lambda_i  \over m^2} + ... \, ]_x \; \phi(x)\nonumber \\
% &=& 0.
%\eea
Also, the fact that  the condition in equation \ref{eq:condition} is covariant, means that for 
{any boost}  $\Lambda$ , 
$
[ D^\Lambda_0 ,   D^{\Lambda i} D^\Lambda_i   ]_x \; \phi(x)
= 0 $, and therefore  
\beq
   [ D^\Lambda_0 ,    D^{\Lambda\mu} D^\Lambda_\mu   ]_x \; \phi(x) =
  [ D^\Lambda_0 ,    D^{\mu} D_\mu   ]_x \; \phi(x) =0 .\eeq
These two facts now imply that  
\beq
\sqrt{D^{2} +m^2-\Lambda_0^\mu D_\mu \Lambda^0 _\nu D^{\nu}}\; 
\Psi({x}^\mu) =i \Lambda_0^\mu D_\mu  \;\Psi({x}^\mu) .
\eeq
Thus,  we conclude that the square-root Schr\"{o}dinger equation is covariant 
 if there  are no charges.   In general, however, we see that 
 the process of reducing and
then quantizing implies the loss of symmetries that were present in the original action.
%%%%%%%%%%%%%%%%%% end 1

\section{Dirac quantization}
\setcounter{footnote}{0}
As described in \cite{Dirac} we begin by quantizing all the variables in the theory,  
$ 
x^\mu \rightarrow \hat{x}^\mu$, $p_\mu \rightarrow \hat{p}_\mu
$, with the commutator algebra $[ \hat{x}^\mu, \hat{p}_\nu ] = i\,
g^\mu_{\:\:\nu}$. 
The  original extended Hilbert space can be described, for example, by 
\beq
I= \int d^4 x \; |x^\mu  \r  \l x^\mu |, \:\:\:\: \l x^\mu | y^\mu\r = \delta^4 (x^\mu -
 y^\mu),  
 \eeq
 and 
\beq
I= \int d^4 p \; |p_\mu  \r  \l p_\mu |, \:\:\:\: \l p_\mu | p_{\mu}' \r = \delta^4
(p_\mu -  p_{\mu}') , 
\eeq
with the commutator algebra   represented by  $| \psi\r \sim \l x^\mu |
\psi\r \equiv \psi(x^\mu)$ and  
$
\hat{x}^\mu \rightarrow x $, $ \hat{p}_\mu \rightarrow -i{\partial
/ \partial x^\mu}$, 
which implies $ \l x^\mu| p_\mu\r = A \exp ( i\, x^\mu p_\mu)$. The physical ``subspace'' is 
 defined by the Dirac condition
\beq
\hat{\Phi} \, |\psi^D\r =0 , 
\eeq
where some ordering for the constraint has been chosen.
The physical states, then, are the solutions to the Klein-Gordon equation (in the context of
quantum gravity   this equation  is  known as the Wheeler-DeWitt equation).
In the physical space, however, we cannot use the    extended state space inner product,
\beq
\l \psi^D | \varphi^D\r = \int d^4 x \, \left( \psi^{D}(x^\mu)\right)^*\, \varphi^D(x^\mu), 
\eeq
because it yields divergent norms for the physical states 
(unless the coordinate space is compact, a case which
will not be considered here). Strictly speaking, physical states are not
in the  extended Hilbert space---the states in the original Hilbert
space had finite norms.
Thus, we need to redefine a   finite  inner product  in the physical subspace, with the right
properties:
 \begin{itemize}
\item[A) ] it satisfies $\left( ( \psi_a |\psi_b ) \right)^*
=                                                     
   ( \psi_b | \psi_a ) $,  

\item[B) ] it is  invariant under changes of gauge-fixing, 

\item[C) ] it preserves the original hermiticity structure, and

\item[E) ] $( \psi | \psi  ) \geq 0$,      $=0$ only if $| \psi  ) =0$.
\end{itemize}
 The last requirement is needed to 
relate amplitudes with 
probabilities. For the case of a single constraint, this inner product is formally given by
\cite{TeitelBook,mythesis} 
\beq  \label{eq:IP}
(\psi^D_a, \psi^D_b) = \int dV  \:   
\left( \psi^{D}_a (x^\mu) \right)^* \; 
\overbrace{  
\delta( {\Upsilon} ) \{ {\Upsilon} ,   {\Phi} \} 
} \;\psi^D_b(x^\mu), 
\eeq 
where $\Upsilon$ is a gauge-fixing function, and the over-brace reminds us that  underneath it there 
is an operator that may need ordering.
As it stands, this is just a recipe that is well-defined only for the case of a
constraint that can be made into a momentum by a canonical transformation\footnote{This can
always  be done  locally, but   in general not  globally.}, $\Phi=
P$. In such a situation the meaning of this expression is rather simple: classically, since a momentum
variable vanishes, say $P$,  its conjugated variable $Q$ is pure gauge and can be ignored. The
quantum version of this is that because of the Dirac condition the  physical states   do  not
depend on the coordinate variable in the coordinate representation, and therefore the inner
product in coordinate space must not include an integration over $Q$.  The determinant insures
that for reasonably chose gauge-fixing functions, 
the effective delta function is precisely $\delta(Q-Q_0)$, for some irrelevant 
constant $Q_0$. Notice that, strictly
speaking, the absolute value of the determinant should be used, and the resulting inner product would
then be positive-definite. 

When the constraint cannot be made into a momentum---as it happens in the 
 relativistic particle case---this recipe cannot be interpreted in such simple terms, and
 is must be taken as  a formal recipe that needs to be
carefully implemented.   We intend to do this now. 

An immediate observation is that the 
operator 
$\overbrace{   \delta( {\Upsilon} )
 \{ {\Upsilon} , {\Phi} \} }$  should be 
hermitean  in the original inner product. 
Hermicity of this operator in the original inner product is essential if the ``reduced''
 inner product is to satisfy $ [ ( \psi_a , \psi_b ) ]^* = ( \psi_b , \psi_a )
$, as we  wrote above, and this will ensure 
  that states have real norms,
\beq 
\para \psi \para^2 \equiv ( \psi  , \psi  ) = [ ( \psi  , \psi  ) ]^* \in \IR 
\eeq
 This is certainly needed if we are to make contact with the classical world
 through  expectation values of hermitean observables.

Although there are other possibilities, we will use  the following  prescription for hermicity,
\beq
\overbrace{   \delta( {\Upsilon} )
 \{ {\Upsilon} , {\Phi}\} } 
\equiv {1\over 2} \left( 
    \delta( \hat{\Upsilon})  \overbrace{
   \{ {\Upsilon }, {\Phi}\}} + {\overbrace{ \{ {\Upsilon} , {\Phi}\} }}^\dagger
\delta( \hat{\Upsilon} ^\dagger)   \right). 
 \eeq

 In the
 case  of  the free relativistic particle, using $\Upsilon=t - f(\tau)$,  
 this simple inner product prescription translates into 
  \beq
(\psi^D_a, \psi^D_b) = \int dV  \: 
\left( \psi^{D}_a\right)^* \;  \left( \delta( t - f(\tau))
\,\hat{p}_t + \hat{p}_t^\dagger\,  \delta( t - 
f(\tau))\right)\;\psi^D_b .
\eeq
Now, in the coordinate representation $\hat{p}_t =  -i \stackrel{\rightarrow}{d/ dt} =  i \stackrel{\leftarrow}{d/ dt} 
=\hat{p}_t^\dagger
$, 
because (the delta function takes care
of any boundary condition problems at $t=\pm \infty$). 
After integrating this delta function, we recover the  
Klein-Gordon inner product, 
\beq 
(\psi^D_a, \psi^D_b) =\left.  \int d^3x \; 
\left( \psi^{D}_a (t,x ) \right)^* \;
  \left(  -i \stackrel{\rightarrow}{d\over dt} + i \stackrel{\leftarrow}{d\over
dt}\right)\;\psi^D_b(t,x ) \right|_{t=f(\tau)} ,
\eeq
which is real, as promised. 
 In the interacting case,  this reasoning leads to the analogous result, with covariant
derivatives
 replacing the regular ones,
 since 
% (recall that $  
%\Phi = (p_\mu -A_\mu) g^{\mu \nu} (p_\nu -A_\nu) -m^2 
%$)
 the Poisson bracket of the constraint with the gauge-fixing term is 
\beq 
\{ t,\Phi \} =  2 g^{0 \nu} (p_\nu -A_\nu) .
\eeq 
The symmetrized inner product measure is then
\beq 
 \delta(\hat{t} - f(\tau)) \hat{g}^{\mu 0} (\hat{p}_\mu - 
\hat{A}_\mu) \, 
  +   \{ \hat{g}^{\mu 0} 
(\hat{p}_\mu - \hat{A}_\mu)\}^\dagger  \, \delta(\hat{t} - f(\tau)),
\eeq 
which leads precisely to the Klein-Gordon inner product in a curved spacetime
\beq (\psi_1, \psi_2 )= \int d^3 x \, \sqrt{|g|} \, g^{0\nu} \, ( \psi^*_1  D_\nu 
\psi_2 - \psi^*_2  D_\nu  \psi_1 ). 
 \eeq
This inner product is related to the $U(1)$ N\"{o}ether current conservation law
of the action $
I=\int d^4 x \, \sqrt{|g|} \, (g^{ab} \,  D_a \psi D_b \psi^* +m^2 \psi \psi^*)
$  
which is  just 
$ 
\nabla_a (\psi D^a \psi^* - \psi^* D^a \psi)=0
$.

 Another  feature of Dirac quantization  is that  observables in the theory are defined
by requiring that they commute with the constraint. For instance, in the case of the parametrized
non-relativistic particle, this means that observables satisfy
the Heisenberg equation of motion with the ``wrong'' sign,
$
i{\partial \hat{O}\over \partial {t}} + [\hat{O},\hat{H}]=0,
$, 
i.e., they are constants of the motion.
The result is that the expectation value of observables between physical states is $\tau$-independent,
and $\tau$ disappears from the formalism.  This, on the other hand, was expected:  in this
formalism ``time'' is a gauge degree of freedom, and everything in Dirac quantization
has been designed to eliminate gauge-dependence.
Recall that this feature was also present in the classical reduced phase space discussion. There it was
 pointed
out that to make  direct contact with the  un-parametrized case the restriction to observables must not
be made.  This is also true here.  However, if  the condition for observables is imposed, 
it is certainly still possible
to recover the concept of time.  For example, in the parametrized non-relativistic case, the observables
associated to position are $\hat{x}_\lambda=\exp (-i\hat{H}(t-t_\lambda)) \hat{x}\exp (i\hat{H}(t-t_\lambda))$, 
and the expectation
value between Dirac states of this quantity depends on $t_\lambda$---and not on gauge-fixing.  The propagator
in this formalism is obtained through the expectation value of eigenstates of  $\hat{x}_\lambda$ and $\hat{x}_{\lambda'}$, and not as the coordinate representation of an evolution operator---there really is no time
or time-evolution.  The same results can be obtained by a trick: to simulate time, make things
gauge-dependent, which is why  parametrized actions are gauge-dependent.
%%%%%%%%%%%%%%%%%%%  end #2 
\section{BRST quantization}
\setcounter{footnote}{0}
In this section we will derive the inner product for the relativistic particle from
the BRST quantization approach.  We will first discuss the correct BRST state
space: this results from a derivation of the  BRST cohomology  using states with
well-defined inner products. Once we have this
state space, we will select its zero ghost  subspace (which consists of two sectors), 
since zero ghost states  are the ones
related to  expectation values of physical observables.  Using these zero ghost states we will then write the 
expectation value of an operator that acts as a propagator, which leads to the BFV
path integral. We  then  argue that the above 
expectation value should be used to define an inner product in the zero ghost sectors, and obtain the inner product 
in each sector.

In BRST quantization the constraints are  automatically implemented together with  the simplectic structure
of the reduced phase space in the theory.  The  Jacobian factors
which must appear in the definitions of brackets or within path integrals, are automatically 
produced by the formalism through the ghosts.

Recall the following from the classical BRST treatment of a system with 
a single
constraint 
$\Phi$ 
\cite{TeitelBook,BFV}.   The first objects that are 
introduced into the 
extended phase space---if not already present---are the multiplier $\lambda$  and its conjugate 
momentum $\pSl$,    
i.e., $\{ \lambda 
, \pSl \} =1$. Since $\lambda $ is an  arbitrary function of $\tau$, its momentum is 
constrained, 
$ \pSl 
\approx 0$. We thus have two constraints. To each constraint, the 
rules say we 
must associate a canonically conjugate pair of ghosts, 
$ \etaSPhi   , \rhoSPhi   $ and $\etaSpSl , \rhoSpSl  $  with $\{  \etaSPhi   , \rhoSPhi   \} 
=1$ and $\{ 
\etaSpSl , \rhoSpSl   \} =1$---and with the other 
(super)brackets vanishing (ghosts are odd Grassmann variables \cite{TeitelBook}, e.g.,  $\etaSPhi^2=0$). 
The total extended phase space is thus 
described 
by $x^\mu, p_\mu , \lambda ,  \pSl ,  \etaSPhi   ,  \rhoSPhi   , \etaSpSl  ,  \rhoSpSl   $.  
The  generator of gauge 
transformations in extended phase space is the BRST generator, 
\beq
\Omega=  \etaSPhi    \Phi  + \etaSpSl   \!\! \pSl, 
\eeq
which has the crucial property   $\{\Omega,  \Omega\} = 0$. This property encodes the constraint algebra. 
The dynamics are 
then 
generated by the hamiltonian ${\cal H}  = h + \{ {\cal O}, \Omega\} $   
(which reduces to $  \{ 
{\cal 
O}, \Omega\} $ when $h=0$), where ${\cal O}$ is a gauge-fixing function. 
In the case of the relativistic particle, one can check that the reduced phase space
equations of motion are correctly produced by the formalism, without having to worry about
the dynamics of the constraints as in the reduced phase space formalism.

In the transition to the quantum theory, phase space coordinates  become operators, 
and the (super)Poisson bracket 
structure is 
translated into the (super)commutator language,  $
\{A,B\} \longrightarrow i\hbar [\hat{A},\hat{B}]
$.  In this case we have both commutators and anti-commutators. For 
example, 
have  \beq
[ \hat{\Omega},\hat{\Omega} ] =\hat{\Omega}^2 =  0 .
\eeq

 In the  parametrized 
relativistic particle case  the state space $   \{|  \Psi  \r\} $ is spanned by the  
basis $ | x^\mu ,\lambda , \etaSPhi   ,\etaSpSl   \r  $, for example.
In the  coordinate   representation we have
\bea
 \l  x^\mu,\lambda , \etaSPhi   ,\etaSpSl    |  \Psi  \r  &\equiv&   \Psi(x^\mu, \lambda, \etaSPhi   ,\etaSpSl )  \nonumber \\
& =& \psi(x^\mu, \lambda)  +\psiZPhi(x^\mu, \lambda)   \etaSPhi      + \psiZpSl(x^\mu, \lambda)   \etaSpSl   + \psiZPhila(x^\mu, \lambda)  \etaSPhi   
\etaSpSl   . 
\eea
%where the  $\psi $'s are functions of $x,t,\lambda $.
The inner product in this  (extended) space is given by
\beq\label{eq:in}
(\Sigma ,\Psi ) \equiv \int {dt dx d\lambda  d \etaSPhi    d\etaSpSl}  \,   
\left(\Sigma  
(z^A)\right)^*\;\Psi (z^A) .
\eeq
The BRST physical space is defined the  BRST generator $\hat{\Omega}$ 
properties:
\begin{quote}
a) $\hat{\Omega}^\dagger =  \hat{\Omega}$, and  \\
b) $\hat{\Omega}^2=0$, 
\end{quote}
which it inherits from the classical description:  $\Omega$ is real, and 
$\{ \Omega, \Omega\}  =0$ \cite{TeitelBook,BFV}.
The definition consists of two items:
 \begin{quote}
i) the BRST physical condition
\beq
\hat{\Omega}  |  \Psi  \r _{Ph}  \equiv 0, 
\eeq
ii) and the concept of  BRST cohomology: we identify
\beq
 |  \Psi  \r _{Ph} \sim   |  \Psi  \r _{Ph} + \hat{\Omega} | \Delta \r , 
\eeq
since the state $\hat{\Omega} | \Delta \r $ is physical 
($\hat{\Omega}^2=0$), and 
has  zero inner 
product with any physical state 
(since $\hat{\Omega}^\dagger 
=  
\hat{\Omega}$,  and  
$\hat{\Omega}  |  \Psi  \r _{Ph}  = 0$). Such states are  called {\em null states}.
\end{quote}
%%%%%%%%%%%%%%%%%%%%%%%%%%%%%%%%
The hamiltonian is given by ${ \hat{\cal H}} = \{ \hat{ \cal O}, 
\hat{\Omega} \}$, 
for 
some gauge-fixing operator $\hat{ \cal O}$, and we might guess   that 
it has no 
effect on physical 
amplitudes
due to the two conditions above. We will see that this is also false.

%%%%%%%%%%%%%%%%%%%%%%%%%
There are many solutions to the BRST condition, $\hat{\Omega}  |  \Psi  \r _{Ph}  \equiv 0$.
Consider for example the zero-ghost number BRST physical states
\bea 
\label{eq:chi}
|\Psi_\Upsilon \rangle &=& |\psi_{\Upsilon =0},  {p}_\lambda \!=\!  \etaSPhi   \!=\! 
\rhoSpSl    \!=\!0\rangle \sim \psiZPhi\hspace{-1.8ex} _{  {p\!}_\lambda \! = \! \Upsilon =\! 0 } \, 
\etaSPhi  , \nonumber \\
|\Psi_\Phi \rangle &=& | \psi_{\Phi= 0} , \lambda \!=\! \rhoSPhi   
\!=\! \etaSpSl  
\!=\! 0
\rangle \sim  \psiZpSl \hspace{-1.4ex}_{ {}_{\tPhi=\!\lambda \! =\! 0} }
\etaSpSl .\eea
We use a notation in which $\psi_{  {p\!}_\lambda \! =\! 0 }$, for example, 
 is shorthand for a state that satisfies $\hat{p}_\lambda \psi =0$.
 These  are the BRST-invariant states which are used in the definition of the boundary conditions
of the BFV path integral, needed in the Fradkin-Vilkovisky theorem 
\cite{TeitelBook,BFV}.
In the coordinate representation,  the BRST equation reads
\bea
\hat{\Omega} \, \Psi &=&
 ( \hetaSPhi \hat{\Phi} +  \hetaSpSl \hat {p}_\lambda )   \, 
\left( \psi 
+ \psiZpSl  \, \etaSpSl  
 + \psiZPhi  \, \etaSPhi   
+\psiZPhila \,\etaSPhi   \etaSpSl  \right)  \nonumber \\
&=&     
\hat{\Phi}\psi \, \etaSPhi  +  \hat {p}_\lambda   \psi  \,  \etaSpSl 
+
\left(\hat{\Phi}  \psiZpSl   -   \hat {p}_\lambda  \psiZPhi   \right)\, \etaSPhi   \etaSpSl
 \nonumber \\ &=& 0 . 
\eea
The  general solution to this equation has the  components $   \psi=\psi_{{}_{\Phi\! =\! p_\lambda=0}}$,
$  \psiZpSl = \hat {p}_\lambda \,\varphi + \!\psiZpSl
\hspace{-1.4ex}_{ {}_{\tPhi \! =\! 0} } $, $\psiZPhi =  \hat{\Phi} \, \varphi +\! \psiZPhi\hspace{-1.8ex} _{  {p\!}_\lambda \! =\! 0 }$ and $\psiZPhila =   { free}$, 
with $\varphi(x^\mu, \lambda)$ arbitrary.   
%\bea \label{eq:solBRST}
%\psi &:& \:\;\;\:    \hat{\Phi}\psi= \hat {p}_\lambda   \psi =0 \\
%\psiZpSl &:&  \:\;\;\:    \psiZpSl = \hat {p}_\lambda \,\varphi + \!\psiZpSl
%\hspace{-1.4ex}_{ {}_{\tPhi \! =\! 0} } \\
%\psiZPhi &:&   \:\;\;\:     \psiZPhi =  \hat{\Phi} \, \varphi +\! \psiZPhi\hspace{-1.8ex} _{  {p\!}_\lambda \! =\! 0 }\\
%\psiZPhila &:& \:\;\;\:    \mbox{ free} .
%\eea
%and we just need to ensure that these states have well defined norms.

%To pin things further down  we need to understand the interplay between the construction
%of the cohomology and the inner product defined above, which we do next.
%
% 
%  
%%%%%%%%%%%%%%%%%%%%%%%%%%%%%%%%%%%%
%
%\subsection{Analysis of the state cohomology} 
  Let us review the
logic in the construction  of the BRST cohomology.  
The standard argument  is 
that 
since\footnote{This is trivially true in this simple case where there is 
no 
place for anomalies to appear in the transition 
$\{\Omega,\Omega\}_{PB}=0 
\longrightarrow [\hat{\Omega}, \hat{\Omega}]=0$}  
$\hat{\Omega}^2 = 0 $ we must make the identification $ |\Psi\r \sim |\Psi\r + 
\hat{\Omega} |\Lambda\r$.  The reasoning is that the states 
$\hat{\Omega} |
\Lambda\r$ are physical (which is clearly true),  and that they decouple from other 
physical 
states.
This, however, is not true in general, because in the physical 
space the inner product is not well-behaved.  That is, the above 
argument is 
based on the assumption that 
$  
\langle \Lambda| \hat{\Omega}^\dagger |\Psi_{\Omega =0}\rangle 
=0,
$ 
and this 
puts restrictions on what $|\Lambda\rangle$ can be, since in general $\hat{\Omega} |\Lambda\r$ is  not null.
The problem can be traced to the following.  In BRST quantization one is eventually  trying
to work with  state spaces defined through a condition similar to $\hat{P}|\Psi\r =0$, which is selecting
something like a zero momentum eigenstate. It is also desired that operators such
as $\hat{P}$ and $\hat{Q}$ be hermitean
in the state space, where the physical states must have finite norms. Finally, the algebra must be preserved,
$[ \hat{Q},\hat{P}]=i$.  It is not too hard to see that it is not possible to satisfy all these conditions
at once. For instance, the quantity $\langle P=0| 
 [\hat{Q},\hat{P}] |P=0\rangle $  is  undefined if all the conditions are met.
 BRST gets around this problem by
the use of different sectors in the state space.   Instead of having a single state space,
there are always pairs of state spaces, and each has its dual, as we will see. By dual of a
sector  we mean
that part of the  state space that is not orthogonal to the sector. This is governed by the
ghost component of each sector: to obtain a nonzero inner product a $\eta_0\eta_1$ combination is needed---see
equation \ref{eq:in}.

 %%%%%%%%%%%%%%%%%%%%%%%%%%%%%%%%%%%%%%%%%%%%%%%%%%%%%%%%%%%%%%%%%%%%%%%%%%%%%%%%%%%%%%%%
 For this reason
it  is incorrect 
to assume, for example,  that one can express a state in the 
form $  
 \Psi_{\Omega\! =\! 0 }  = 
  \psi_{{}_{\tPhi\!= \!{p\!}_\lambda \! =\! 0 }} + \psiZpSl \hspace{-1.4ex}_{
  {}_{{}_{\tPhi\!= \!{p\!}_\lambda \! =\! 0 }} } \etaSpSl    + \psiZPhi\hspace{-1.8ex} _{ {}_{{}_{\tPhi\!= \!{p\!}_\lambda \! =\! 0 }} }
 \, \etaSPhi   +\psiZPhila\hspace{-2.7ex}_{ {}_{{}_{\tPhi\!= \!{p\!}_\lambda \! =\! 0 }} } \,\etaSPhi   \etaSpSl  
 $ by a proper choice of $|\Lambda\rangle$, since these states have an ill-defined norm.  
%%%%%%%%%%%%%%%%%%%%%%%%%%%%%%%%%%%%%%%%%%%%%%%%%%%%%%%%%%%%%%%%%%%%%%%%%%%%%%%%
However, because all the states in the physical 
space must have finite norms,  it is possible, by the addition of a null state, to rewrite this solution
in the form 
\beq
\label{eq:coho} 
 \Psi_{\Omega\! =\!0 }  = 
   \varrho_{{}_{\Phi\!= \!{p\!}_\lambda \! =\! 0 }} +
\varrho_{{}_{\Phi\!= \! \lambda \! =\! 0 }}^1
% \psiZpSl \hspace{-1.4ex}_{ {}_{\tPhi \!=\!\lambda\! =\! 0} } 
\etaSpSl  
 +  \varrho_{{}_{\Upsilon\!= \! \lambda \! =\! 0 }}^0
% \psiZPhi\hspace{-1.8ex} _{  {p\!}_\lambda \! =\! \varphi \!=\!0 }
 \, \etaSPhi   +
+  \varrho_{{}_{\Upsilon\!= \! \lambda=\! 0 }}^{01}
 % \psiZPhila\hspace{-3.4ex}_{ {}_{\varphi \!=\lambda\! =\! 0 } }  \,\etaSPhi   \etaSpSl    .
\eeq   
where $\Upsilon=\Upsilon(t)$ is some canonical gauge-fixing, and $\varrho=\varrho(x^\mu)$.
Notice that this  state has a well defined norm, as it should, and that 
in each sector a constraint and a gauge-fixing term are  satisfied, while
in the dual sectors the roles of the multiplier and original constraint are reversed. 
 Because of this, these states have finite norms. This components include as particular cases
 the states in equation \ref{eq:chi}---the boundary states
in the BFV path integral.
%{ Notice also that duality of the sectors is lost if the constraint has more than one branch.}

Let us summarize the above discussion with the following statement:
let   ${\cal F}$ be the space of states of finite norm; 
any BRST-invariant state in ${\cal F}$ can be brought into the   form of equation \ref{eq:coho} by the addition
of a null state in ${\cal F}$.

\subsection{Inner product in the zero ghost sector}
\label{sec:zeroghost}

 Although we have an inner product in the physical state space,  we would 
 like to avoid the use of dual state spaces, and define the theory within a single 
 state sector  with a unique Hilbert space. the zero-ghost sectors are
  associated with the physical observables---which have zero ghost number---so they are
  the natural candidates for this task.
 To achieve this goal we need an operator that will map a state in a given sector to one in the dual
 sector. The operator $\exp ( [ \hat{K}, \hat{\Omega}])$ has the right characteristics,
 as we  now explain.   Any definition of inner product in BRST theory must
take into account the existence of the BFV path integral 
in phase space \cite{TeitelBook,BFV}.
The path integral carries information about the states, the inner product, and the
hamiltonian. As we discuss below, it is  easy to 
 write  the physical amplitude in terms of a path integral in phase space. The concept
of physical state space  enters in the boundary conditions, which arise from the required BRST invariance of
the ``end'' states in the amplitude.  This provides the interpretation for the usual  BRST boundary 
conditions:
they can be understood in the context of the cohomology of BRST-invariant states.
 States
that implement such boundary conditions belong to the cohomology of the 
BRST generator.
%%%%%%%%%%%%%%%%%%%%%%%%%%%%
For instance, in the 
relativistic particle case   we can derive the path integral  expression for the 
propagation amplitude
from the Hilbert space expression  \cite{mythesis}\beq  
U(t_i,x_i,t_f,x_f)\equiv \langle t_f,x_f,\etaSPhi \!=\! \rhoSpSl \!=\!  {p}_\lambda \!=\!0|\; 
{\hat{\cal U}} \;
|
t_i,x_i,\etaSPhi \!=\!\rhoSpSl\!=\!  {p}_\lambda \!=\! 0\rangle\eeq   
 by using the states in the zero-ghost cohomology sectors and the 
propagation
operator 
\beq 
{\hat{\cal U}} = e^{-i \Delta\tau\hat{\cal H}}
\eeq
  where  $\hat{\cal H}$ is the 
extended
super-hamiltonian: $ {\hat{\cal H}} \equiv  \{ \hat{O}, \hat{\Omega}\}$. 
Our notation anticipates that this amplitude will not
depend on $\Delta \tau$, which will be the case. To obtain the path integral expression
rewrite
 ($-i \Delta\tau\hat{\cal H} =   \Delta\tau  \, [ {\hat{\cal O}}  , \hat{\Omega}] $)
\beq
e^{ \Delta\tau  \, [ {\hat{\cal O}}  , \hat{\Omega} 
]} = \lim_{N\rightarrow \infty} \left( 1+{ \Delta\tau  \over N} [  \hat{\cal O}  
, \hat{\Omega} 
]\right) ^N \eeq 
 and, at each $\tau$-division, insert the resolution of the identity in extended phase space
\bea  
\hat{I}  &=& \int dtdxd p_\lambda d\etaSPhi d\rhoSpSl\: |t,x,  {p}_\lambda , \etaSPhi
,\rhoSpSl\rangle \langle t,x,  {p}_\lambda ,\etaSPhi ,\rhoSpSl| \nonumber \\
&=&
 \int dp_tdp_xd\lambda d{ \rhoSPhi}d{\etaSpSl} \:
|p_t,p_x,\lambda,{ \rhoSPhi}, {\etaSpSl}\rangle\langle 
p_t,p_x,\lambda,{\rhoSPhi},{\etaSpSl}|, 
\eea 
and  the projections 
\beq 
\langle t,x,  {p}_\lambda ,\etaSPhi ,\rhoSpSl|p_t,p_x,\lambda, {\rhoSPhi},{\etaSpSl}\rangle 
=
e^{i\left( tp_t+xp_x+  {p}_\lambda \lambda+\etaSPhi {\rhoSPhi}+\rhoSpSl{\etaSpSl} \right)} , 
\eeq 
just as in the unconstrained case.
  
To be more specific, let us consider the following two types of gauge-fixing terms:
\begin{itemize}
\item[a) ] {\bf Non-Canonical:} ${\cal O}_{NC} =    \rhoSPhi   \lambda$,  which 
yields
 \beq \{{\cal O}_{NC} ,\Omega\} = 
\lambda \Phi +  \rhoSPhi  \etaSpSl . \eeq 
The terminology refers to the fact that 
the gauge-fixing is carried through the lagrange multiplier.
\item[b) ] {\bf Canonical:} ${\cal O}_{C} =  \rhoSpSl   \Upsilon $, where 
$\Upsilon = t - f(\tau) $. Here
 \beq 
\{ {\cal O}_{C} ,\Omega\}=  \rhoSpSl    \etaSPhi   \{\Upsilon, \Phi\} +
 \pSl \Upsilon ,
 \eeq 
and the resulting gauge-fixing is essentially $\Upsilon=0$.
\end{itemize}
Notice that the sum of these  two terms  is an ``anti''-BRST charge, $
{\cal O}_{NC} +{\cal O}_{C}=\rhoSPhi   \lambda + \rhoSpSl   \Upsilon $, the result of replacing 
each coordinate in the BRST generator by its phase space conjugate---with the exception of 
the quadratic constraint, which has as conjugate the gauge-fixing term.
Now, formally
\beq  \langle {\Psi_a}|e^{ [\hat{{\cal O}} ,\hat{\Omega}]}\;  | {\Psi_b} 
\rangle= 
\langle {\Psi_a}| {\Psi_b} \rangle\eeq for physical              
states---states
annihilated by the BRST generator---and
this is 
where the
invariance of the amplitude under changes in gauge-fixing is 
expected to 
come  from.
However, this is a formal statement, because the inner product needs regularization.
 For 
example, take 
the BRST invariant states
$|\Psi_{\Phi}\rangle$ of  equation \ref{eq:chi}. Then
$ 
\langle \Psi_\Phi | e^{ [\hat{{\cal O}}_C ,\hat{\Omega}]}\;  |\Psi'_\Phi 
\rangle =$
\beq
 \langle \psi_{{}_{\tPhi=0}}, \lambda\!=\! 
 \rhoSPhi  \!=\!\etaSpSl \!\!\!=\!0|
\exp \! \left( 
\left(\hrhoSpSl \! \hetaSPhi  [\hat{\Upsilon},\hat{\Phi}]+i\hpSl 
\hat{\Upsilon} 
\right)  \right) 
|\psi'_{{}_{\tPhi=0}}, \lambda\!=\!  \rhoSPhi   \!=\! \etaSpSl \!\! \!=\! 0 \rangle, 
\eeq  which, up to
ordering questions,  we can guess will turn out to be
$ \label{eq:KG} \langle \psi_{\Phi=0}|\overbrace{ \{ {\Upsilon},  {\Phi}\} 
\delta( {\Upsilon}) } \; 
|\psi'_{\Phi=0}\rangle
$ 
i.e., the Klein-Gordon inner product. Why? The ghost integrations should just yield the
determinant $\overbrace{ \{ {\Upsilon},  {\Phi}\} }$, and the multiplier degree of freedom the delta
function. Ordering questions are  important, however,  so let us proceed carefully.
The amplitude is 
$$
\l  \psi_a (t,x) ,  \, \Phi \! =\! 0\! =\! \lambda \! =\! \etaSpSl \!\! =\!  
 \rhoSPhi   | 
 \,  \exp
 \left(
 \hrhoSpSl \!\hetaSPhi  [\hat{\Upsilon}, \hat{\Phi}] + i 
\hpSl \hat{ \Upsilon } 
\right) 
|\psi_b (t, x) ,  \, \Phi \! =\! 0\! =\! \lambda \! 
=\! \etaSpSl \!  \! =\!   \rhoSPhi   \r $$ \beq 
 = \int d^4 x \:  \psi_a^*  (t, x)  \left[ \int d  \!\pSl d \rhoSpSl   \!d \etaSPhi   
\; \exp\left({  \rhoSpSl  \etaSPhi [\hat{\Upsilon}, \hat{\Phi}] + i   \pSl \hat{ \Upsilon } 
}\right) \right] \psi_b (t, x) , 
\eeq
where in the free case $ [\hat{\Upsilon}, \hat{\Phi}] = [\hat{t}, \hat{p}^2]= i2 \hat{p}_t $.
To compute this integral we make use of the 
  Campbell-Baker-Hausdorff 
theorem (see, for example, reference \cite{Miller}). That is 
\beq
\ln (e^A e^B) = A+B+{1\over 2} [A,B] + {1\over 12} [A,[A,B]] - {1\over 
12} 
[B,[B,A]] + ...=\eeq 
Fortunately  the series ends quickly here, and we have
\bea
\exp\left(  \rhoSpSl  \etaSPhi [\hat{\Upsilon}, \hat{\Phi}] + i   \pSl \hat{ \Upsilon } \right)
&=& \exp\left(    \rhoSpSl  \etaSPhi [\hat{\Upsilon}, \hat{\Phi}]  - i  \rhoSpSl    \etaSPhi   
\pSl    \right)\; 
\exp\left(   i   \pSl \hat{ \Upsilon } \right) \nonumber \\
 &=&   \left( 1+   
\rhoSpSl  \etaSPhi [\hat{\Upsilon}, \hat{\Phi}]  - i  \rhoSpSl    \etaSPhi    \pSl   
  \right)
 \exp\left(    i   \pSl \hat{ \Upsilon } \right). 
\eea
The ghost integrations are now easily done, and yield
\beq
\int d \!  \pSl d \rhoSpSl   d \etaSPhi   \; 
\exp\left( \rhoSpSl  \etaSPhi [\hat{\Upsilon}, \hat{\Phi}] + i   \pSl \hat{ \Upsilon } \right) =
i \hat{p}_t \delta(\hat{\Upsilon}) + i  \delta(\hat{\Upsilon}) \hat{p}_t . 
\eeq 
This is, up to a factor of $i$,   the Klein-Gordon inner 
product, since
 \beq 
\l \psi_a | \delta(\hat{t} - f(\tau)) \hat{p}_t |\psi_b\r = 
\int dx dt\, \l \psi_a | x,t \r \l x,t | \delta(\hat{t}\! -\! f(\tau))
 \hat{p}_t |\psi_b\r
 \eeq
 which is just 
\beq \int dx dt\, \delta( t\! -\! 
f(\tau)) \, \l \psi_a | x,t\r \,\left( -i{\partial \over
\partial t}\right) \,  \l x, t|\psi_b\r 
\eeq   
{Notice that this inner product  involves the hermitized expression of the
determinant  operator---as discussed
in the previous section.  It does not yield an absolute value or anything else.}
 
 With  an
electro-magnetic background   the result is again immediate:
\beq
\int d \!  \pSl d \rhoSpSl   d \etaSPhi   \; 
\exp\left(  \rhoSpSl  \etaSPhi [\hat{\Upsilon}, \hat{\Phi}] + i   \pSl \hat{ \Upsilon } \right) =
i \hat{\Pi}_0\, \delta(\hat{\Upsilon}) + i  \delta(\hat{\Upsilon}) \, \hat{\Pi}_0
\eeq
because $
[\hat{t},\; \hat{\Phi}_E ] = 2i \hat{\Pi}_0$ , and $[\hat{t} ,\; [\hat{t}, \hat{\Phi}_E ] ] = -2$,
which is again a a multiple of the identity and commutes, and the series terminates. 
Is this true for the general interacting case?  
We can compute 
 \beq 
 [ \hat{t},\,  \hat{\Phi} ] = 2i \, \hat{g}^{\mu 0} ( \hat{p}_\mu-
 \hat{A }_\mu)+\nabla_\nu  \, \hat{g}^{0 \nu} =  2i  \,  \hat{g}^{\mu  0}  ( \hat{p} , 
 _\mu- \hat{A}_\mu) 
 \eeq 
 and  $  [\hat{t} ,\;    [\hat{t}, \;  \hat{\Phi} ]]
 =- 2 \hat{g}^{00} $, and again we need   $ [ \hat{g}^{00} ,\; 
 \hat{g}^{\mu  0}  ( \hat{p}_\mu- \hat{A}_\mu)   ] =0  $. This holds
provided $ \hat{g}^{0\nu} [\hat{p}_\nu ,\; \hat{g}^{00}]=0$.
% This can  be satisfied by
%the proper choice of a coordinate system.
 Notice that a natural (covariant) ordering of the constraint operator is
 assumed.

We should  also study the other zero-ghost sector,  with the states  $| \Psi_\Upsilon\rangle$, which satisfy
        $
\hpSl,\;\hat{\Upsilon},\; \hetaSPhi ,\; 
\hrhoSpSl    \; \: |
\psiZPhi\hspace{-1.8ex} _{\Upsilon \! =\!   {p}_\lambda \! =\!  \eta_{\,\tPhi} =\! 0  }
 \r
=0$.  These are not Dirac states, and appear in
 the path integrals we   evaluate  in the particle case \cite{Gomis,BFV}.
It is easy to see, as we now show, that using the non-canonical gauge-fixing, the 
amplitude (from which it is again easy to reach the  BFV path integral) becomes
\beq 
\langle \Psi_\Upsilon| 
e^{ [\hat{{\cal O}}_{NC} ,\hat{\Omega}]}\; 
|\Psi'_\Upsilon \rangle
= \langle \psi_{\Upsilon=0}|\delta(\hat{\Phi})\;  |\psi'_{\Upsilon =0} \rangle,
\eeq 
i.e.,  the Hadamard, or 
on-shell, amplitude:
\bea  
 (\psi,\psi') &\equiv & \l   \psi _{\Upsilon \! =\!   {p}_\lambda \! =\! 
\eta_{\,\tPhi} =\! 0  }|  \; e^{ [\hat{\cal O}_{NC},\hat{\Omega} ]}\;
         |  \psi'  _{\Upsilon \! =\!   {p}_\lambda \! =\!  \eta_{\,\tPhi} =\! 0  }  \r 
\nonumber \\
&=&
\int d \etaSPhi   d\etaSpSl \:
 \etaSPhi   \etaSpSl 
 \l   \psi_{\Upsilon \! =\!   {p}_\lambda \! =\! 0 }| 
\; e^{ [\hat{K},\hat{\Omega} ]}\;
         |  \psi'  _{\Upsilon \! =\!   {p}_\lambda \! =\! 0 } \r
 \nonumber \nonumber \\ &=&
\int d \etaSPhi   d\etaSpSl \etaSPhi   \etaSpSl  \l  \psi _{\Upsilon \! =\!   {p}_\lambda \! =\! 0 }  | 
\; \exp \left( {i \hat{\lambda}\hat{\Phi} + \hrhoSPhi    \hetaSpSl  } \right)
 | \psi'_{\Upsilon \! =\!   {p}_\lambda \! =\! 0 } \r 
  \nonumber\\
&=& \label{eq:BFVDIRAC} \l \psi _{\Upsilon\! =\! 0} | \delta(\hat{\Phi}) |
\psi'  _{\Upsilon\! =\! 0} \r .
\eea
Notice that the full-range of the multiplier $\lambda$ has been used, since in this representation
this variable is fully ranged. 
This is a quite remarkable result: neither the initial nor the final states are physical in the
Dirac sense, and this object has no natural interpretation as a physical amplitude  within the
Dirac approach.  It appears
naturally in the BRST approach.  We will next see how this amplitude appears in (BRST)-Fock quantization
as well.  For now, notice that this inner product has a lot of nice properties (as long as $\Phi$ is 
hermitean), including positivity.

 %%%%%%%%%%%%%%%%%%%%%%%%%%%%%%%%%%%%%%%%%%%%%%%%%
\section{Dirac-Fock quantization} \setcounter{footnote}{0} 

We  review first the situation for simple constraints \cite{TeitelBook}, i.e., 
constraints can be canonically transformed to 
momenta,  $P_1\approx 0 
\approx P_2$. 
The reason is that this 
approach is really unknown territory---the prescription is clear 
only for the simple class of situations in which the constraints can be made into
momenta by  canonical transformations.   
%%%%%%%%%%%%%%%%%%%%%%%%%%%
 To begin with, 
an even number of constraints is required.  Let us 
state why immediately: we will, in essence,
 pair the constraints and assign to each 
  combination opposite sign norm states so that their effect in 
the theory cancels after
we select the physical space.
Thus, gauge--degrees of freedom effectively disappear from the theory. 
The key new ingredient
here is the appearance of  states with negative norms.

  Fock quantization begins with the definition of  the 
operators
\beq \label{eq:oscillators1}
\hat{a} = \hat{P}_1 + i\hat{P}_2 , \:\: \hat{a}^\dagger = \hat{P}_1 
-i\hat{P}_2,
\eeq 
and
\beq \label{eq:oscillators2} 
\hat{b}=-{i\over 2}(\hat{Q}^1 +i\hat{Q}^2), 
\:\: \hat{b}^\dagger = {i\over 2}(\hat{Q}^1 -i\hat{Q}^2) .
\eeq
The commutation relation that follow from these definitions 
are
\beq
[\hat{a},\hat{b}^\dagger] 
= 
[\hat{b},\hat{a}^\dagger]  = 1
\eeq
and the rest zero.
{Notice that it is implied by the notation here that both 
$\hat{P}_1$ and 
$\hat{P}_2$ are 
hermitean.} For example, $\hat{a}+\hat{a}^\dagger$ is hermitean, and 
is equal  
to $2\hat{P}_1$. This fact is crucial for the development of the formalism, 
and is 
a subtle assumption---it selects an indefinite inner product when we 
define 
the vacuum.

The states on this space are defined by the following construction:  
\begin{itemize}
\item[a) ] It is assumed that there is a ``vacuum'' state,   $|0\rangle$, 
satisfying the 
conditions 
$$
\hat{a}|0\rangle  = \hat{b}|0\rangle = 0 $$

\item[b) ] This state is also assumed to 
have unit 
norm, $\langle 0 |0\rangle = 1$.

\item[c) ] The rest of these states are defined by acting on the ``vacuum'' 
above with 
the creation operators.
\end{itemize}
Before we work out some of the consequences of this prescription, let 
us look 
at what it is doing in terms of the $\hat{P}_i$ operators. What is the 
vacuum? We 
need a state that satisfies 
\beq
(\hat{P}_1 +i\hat{P}_2)|0\rangle = 
-{i\over 2}(\hat{Q}^1 +i\hat{Q}^2)|0\rangle = 0
\eeq
In a standard Hilbert space  
there is no 
such state! The operators $P_1,P_2,Q^1,Q^2$ are assumed to be hermitean, yet if we look at the definition of the vacuum we notice that at some of them need to have imaginary eigenvalues.  How can this be?  It is not to hard to see that an indefinite inner product is needed. 
For example \cite{TeitelBook}, consider the states $(\hat{a}^\dagger +
\hat{b}^\dagger) |0\rangle $,  $(\hat{a}^\dagger - \hat{b}^\dagger) |0\rangle $, and $ 
\hat{a}^\dagger  
|0\rangle $.
 They have positive, negative, and zero norm respectively.  The Fock 
Hilbert space 
is not a positive definite inner product space.

Next, we need to impose the condition for physical states. 
The constraints above are equivalent 
classically to 
demanding that $a\approx 0 \approx a^\dagger$. However, we 
cannot 
demand this condition from the states:  there is no state in our 
construction satisfying  $\hat{a}^\dagger |\psi\rangle=0$!
We can only demand that the physical states satisfy 
\beq 
\hat{a} |\psi_F 
\rangle = 0.
\eeq
These states have a well defined norm: 
{the physical space is a true subspace of the original Hilbert
space}. We do not need to redefine the inner product.  Thus, Fock quantization is superior 
to Dirac quantization in this respect: in Dirac quantization
one needs to redefine an inner product in the physical subspace.  Also, there is no 
need for gauge-fixing.
 
What happened to  the other ``half'' of the constraint? 
Although the physical states 
do not 
satisfy the Dirac condition, the expectation value of the constraints 
between physical states is always zero.   Moreover, we will now see 
that the physical Fock state space is reduced to a space
isomorphic to the  Dirac      
states---the space of states that satisfy the constraints in Dirac      
quantization, which in this simple example  are given by the 
{\em unique} state, $|P_1 \!=\!P_2
\!=\! 0 \r$.  This is because
a physical state  in the Fock space is   either   the vacuum or  a 
linear 
combination of the vacuum and a physical state that has zero inner 
product 
with 
all 
the physical states.
Indeed, the vacuum is a physical state and the other physical states are given
 by  the so-called {\em  null states}, 
$
 \hat{a}^{\dagger n }  |0\r
$, since $[\hat{a}, \hat{a}^\dagger]=0$.
  These decouple from the physical states, since 
$\hat{a}^\dagger\equiv (\hat{a} 
)^\dagger$.
 Null states are, by definition, physical states with zero norm and
zero inner product with any other physical state. 
Thus,  we are left over with a single state, 
just as in the Dirac approach. What has
happened is that the two degrees of freedom have combined 
and have annihilated each other.
This is reflected by the appearance of the null states, 
which in turn is a consequence of the
fact that we have quantized  the system with an 
indefinite inner product (a change of
variables will decouple the algebra  into a commutator with the ``wrong'' sign,  $ [\hat{{\cal
Q}}_- , \hat{{\cal Q}}_- ^\dagger] = -1$,  and a regular one $[ \hat{{\cal Q}}_+ , \hat{{\cal
Q}}_+ ^\dagger] = +1 $,  \cite{mythesis}). 

How can we generalize the above formalism to  more complicated
situations?  Let us study the case of 
 the relativistic particle. 
To proceed with Fock quantization we need two constraints:  the original 
 constraint,
$\Phi\approx 0$, and another one  $P_\lambda \approx 0$. Thus, the form of the action which
includes the lagrange multiplier
 is a more natural starting point for Fock quantization. 
The full state space will be defined  using 
\bea
\hat{a}&=&\hat{p}_t + i \hat{p}_\lambda \nonumber\\
\hat{b}&=&-{i\over 2}(\hat{t}+ i \hat{\lambda})
\eea
and hermitean conjugates, and the usual definition of the vacuum. As long as we use canonical pairs in this
definitions, we will obtain the right commutation properties of the creation and annihilation operator. For example, this definition could be changed using a 
canonical transformation.

The  strategy  will be as follows: \begin{itemize}
\item[I) ] Find an  
operator   $\hat{M}$ 
such that we can write  the constraint as  a linear combination of this
operator and its hermitean conjugate,  for instance
\beq
\hat{\Phi}= \hat{M} + \hat{M}^\dagger .% , \:\:\;\;\:\: i\hat{P}_2= \hat{M} - \hat{M}^\dagger .
\eeq  
Something similar should happen to the other constraint, 
$P_\lambda \approx 0$.
This we will call a {\em splitting}\footnote{It is actually 
sufficient to be able to
express 
the original constraint as a linear combination of $M$ and $M^\dagger$.  See
below.}. 
\item[II) ] Define the physical states by $\hat{M} \, |\Psi \r =0$.
 This will ensure that the expectation value of the constraints in the physical space is
zero,
 $ \ _{Ph}\!\l \Psi |  \hat{\Phi} \, | \Psi \r _{Ph} =  \ _{Ph}\!\l \Psi |  \hat{P}_2 \, | \Psi
\r _{Ph} =0$.

\item[III) ] The splitting must be such that $\hat{M}$ be a {\em weakly normal} operator, 
$[\hat{M}, \hat{M}^\dagger]=\hat{u}\, \hat{M}$. 
This insures that the states generated by
$\hat{M}^\dagger$ are null (i.e., physical states that have  zero
inner product with any physical state).
 \end{itemize}
 %%%%%%%%%%%%%%%%%%%%%%%%%%%%
Let us now consider the  simple quadratic constraint $\hat{\Phi} = \hat{P}_1^2 - \hat{A}^2$,
where  we assume that $\hat{A}$  is a hermitean  operator that commutes with $\hat{a},
\hat{b}$---in the particle case this corresponds to an  electro-magnetic background with zero
electric field, for example: $P_1 = p_t$ and $A^2 = \Pi_i^2 +m^2$.  Now 
rewrite the constraint in terms of the new variables,
\beq
  \hat{\Phi} = \hat{P}_1^2 -  \hat{A}^2 = ( { \hat{a} + \hat{a}^\dagger \over 2})^2 - \hat{A}^2 = 
{1\over 4} \left( \hat{a}^2 + \hat{a} ^{\dagger 2}  + 2 \hat{a} \hat{a}^\dagger \right)  - \hat{A}^2 .
\eeq
  A natural first ``splitting'' guess would be   
 $
\hat{M} = {1\over 4} \left( \hat{a}^2 +   
 \hat{a} \hat{a}^\dagger  \right)  - {1\over 2}  \hat{A}^2
$. 
Notice, however, that this definition  implies that
$
\ _{Ph}\!\l \Psi |   (\hat{M} - \hat{M}^\dagger) \; | \Psi \r _{Ph} =0
$, which translates into
\beq
\ _{Ph}\!\l \Psi |   
{1\over 4} \left( \hat{a}^2 -\hat{a} ^{\dagger 2} \right)  
 | \Psi \r _{Ph} =\ _{Ph}\!\l \Psi |   
i \hat{P}_1 \hat{P}_2
 | \Psi \r _{Ph}  =0, 
\eeq
which is not quite what we need. So how do we perform the ``splitting'' of the
constraint? Although it is not too hard to see the solution, an elegant  solution
 to the above conditions comes out directly form  BRST-Fock quantization, as we discuss next.   
%%%%%%%%%%%%%%%%%%%%%%%%%%%%%%%%%%%%%%%%%%%%%%%%%%%%%%%%%%%%%%%%%%%%%%%%%%%%%%%%%%%%%%%%%%%%%
%%%%%%%%%%%%%%%%%%%%%%%%%%%%%%%%%%%%%%%%%%%%%%%%%%%%%%%%%%%%%%%%%%%%%%%%%%%%%%%%%%%%%%%%%%%%
\section{BRST-Fock quantization}
As  in regular BRST, in  the  BRST-Fock approach we define the physical space by $
\hat{\Omega} |\Psi\r = 0$, with $\hat{\Omega} = \hat{\eta}_0 \hat{\Phi} + \hat{\eta}_1\hat{P}_2 
$.
Now, in BRST-Fock we  define oscillator variables for the ghosts 
\beq
\hat{c}= \hat{\eta}_0 + i\hat{\eta}_1, \:\:\:\: \hat{\bar{c}}= {i\over 2} (\hat{\rho}_0 +i \hat{\rho}_1), \:\:\:\: [\hat{c}, \hat{\bar{c}}^\dagger]=1,
\eeq
and hermitean conjugates, as well as those in equations \ref{eq:oscillators1} and \ref{eq:oscillators2}.
There are other ways to define these variables, but this will not alter the results below in
a fundamental way. 
The states are constructed in terms of a unit-norm vacuum $|0\r$ defined by 
\beq
\hat{a},\hat{b},\hat{c}, \hat{ \bar{c}} |0\r =0.
\eeq
 Using these variables the BRST generator becomes
\beq
\Omega = {\hat{c}+\hat{c}^\dagger \over 2} \hat{\Phi} + {\hat{c}-\hat{c}^\dagger\over 2i}{\hat{a}
-\hat{a}^\dagger \over 2i}=
{1\over 2} \left(\hat{c}\,  \hat{M}^\dagger  + \hat{c}^\dagger \,  \hat{M} \right),
\eeq
where $\hat{M}= (\hat{\Phi} +{\hat{a}-\hat{a}^\dagger \over 2})/2  = (\hat{\Phi} + i \hat{P}_\lambda)/2$. 
Hence, the physical states satisfy
\beq
\hat{c}|\Psi\r = 0 , \:\:\:\: \:\: \hat{M} |\Psi\r =0 .
\eeq
This ensures that $\hat{\Omega} |\Psi\r = 0 $, and  that
 physical states carry no ghost excitations. Notice that  this procedure fixes the ``splitting'' defined
in the previous section, and satisfies
the properties 
\beq
\hat{M}+\hat{M}^\dagger = \hat{\Phi}, \:\;\:\; \hat{M}-\hat{M}^\dagger = i\hat{P}_2 , \:\;\:\; 
[\hat{M},\hat{M}^\dagger]=0 .
\eeq
It is possible to define the ghost oscillator variables  differently, and find that the
states satisfy $\hat{M}' \psi=0$, with $\hat{M}'= \alpha\hat{\Phi}+\beta\hat{p}_\lambda$.  
This ambiguity is not serious, and corresponds to canonical transformation of the ghosts.
 %%%%%%%%%%%%%%%%%
 \subsection{The Fock representation in coordinate space}
 Consider 
the mixed  coordinate-momentum 
representation in which we write the Fock states in the form $\varphi = 
\varphi(p_t, \lambda)$.   The vacuum is defined by 
\beq
 ( \hat{t}  + i\hat{\lambda} ) 
\varphi_0 (p_t, \lambda) =( i{\partial \over \partial  p_t}  + i \lambda) 
\varphi_0 (p_t, \lambda) = 0, \eeq  and 
\beq
( \hat{p}_t + i\hat{p} _\lambda) 
\varphi_0 (p_t, \lambda) =( p_t  + {\partial \over \partial \lambda}) 
\varphi_0 (p_t, \lambda) = 0.
\eeq
This is solved by
\beq
 \varphi_0 (p_t, \lambda) = e^{- 
\lambda p_t}.\eeq 
Can we come up with an inner product here that respects the 
algebra, the hermiticity properties of the operators and that gives 
unit norm to the vacuum, i.e., can we rewrite the Fock inner product in this
representation?  The answer is yes, of course. The inner 
product is given by 
\beq
 (\psi_a, \psi_b ) \equiv \int_ {-i\infty}^{i\infty} d\lambda \int_{-
\infty}^{ \infty}dp_t\, [\psi_a(p_t^*, \lambda^*)]^* 
\psi_b (p_t, \lambda).
\eeq
First notice that the vacuum is normalizable to unity: 
\beq
\dis \parallel
\psi_0 \parallel = \int_{-i\infty}^{i\infty} d\lambda \int_{-\infty}^{ \infty}dp_t\, e^{-2 \lambda p_t} = 
 \int_ { -\infty}^{ \infty} idq \int_{-\infty }^{ \infty }dp e^{-
2iqp } = i\pi. 
\eeq
Next, we can  check that the operators are hermitean. This 
could be troublesome, for example, for the operator 
$ \hat{t} 
\sim 
i\partial_{p_t} $. But we can check  that the boundary term  that appears in the check for hermicity vanishes,
\beq
\int_{-i\infty}^{i\infty} d\lambda \int_{-\infty}^{ \infty}dp_t\,  i\partial_{p_t} e^{-2 \lambda p_t}
= i \pi \int_ { -\infty}^{ \infty} dp_t
\, \delta'(p_t)=0. 
\eeq
%vanishes,
%\beq
% \dis = -\int_ { -\infty}^{ \infty} dq \int_{-\infty }^{ \infty 
%}dp\partial_p e^{-2iqp }= i \pi \int_ { -\infty}^{ \infty} dp_t
%\, \delta'(p_t)=0.
%\eeq
Similarly, it is also easily checked that $\hat{p}_\lambda$, $\hat{\lambda}$, and $\hat{p}_t$ are
also hermitean in this inner product, as they should. This is sufficient to
insure that we have a good representation of the Fock algebra, since the whole representation 
is based on the unit norm of the vacuum, the algebra,
 and the hermicity properties of the operators.

Now, as we discussed, the  condition for physical states in  (BRST-) Fock quantization 
is  \beq
(\hat{\Phi} +i \hat{p}_\lambda ) \, \psi = 0.
\eeq
 In the above representation (and in the free case) this has the
general solution 
\beq
\psi(p_t, \pi) = \varphi(p_\mu)\, e^{- 
\lambda (p_\mu p^\mu - m^2) }.
\eeq
More generally, the solution is given by  $|\Psi_{ph}\r = \exp(-\hat{\lambda}\hat{\Phi}) |\varphi(x^\mu)\r$. 
Observe that the inner product between two such states is  ``on-shell'',
\bea
( \Psi_a|\Psi_b) &=&\int d^4p \,  (\Psi_a(p_\mu))^* \, \delta(\Phi)\,  \Psi_b(p_\mu) \nonumber\\
                   &=&\int d^4x \, (\Psi_a(x^\mu))^* \, \delta(\hat{\Phi})\,  \Psi_b(x^\mu) \nonumber\\
                   &=& \l \Psi_a|\delta(\hat{\Phi})| \Psi_b\r.
\eea
As long as $\Phi$ is hermitean, this inner product satisfies the property $( \Psi_a|\Psi_b)^* =( \Psi_b|\Psi_a)$,  and it never yields negative norms. How can we interpret this result? In the case of the parametrized non-relativistic
particle, where $\Phi=p_t+R(t,x,p_x)$, the interpretation is simple: because of the 
delta function in the definition of inner product, the state space is reduced to the 
 states  of non-relativistic quantum mechanics
$
\Psi_a(x^\mu)= ( x^\mu_r|\Psi)$ ,  with  $ |x^\mu_r) = \exp({i\hat{R}t}) |\vec{x})$.
To see why, notice that null states are states of the form
$
\varphi_{null}= f(\hat{\Phi}) e^{- \lambda \hat{\Phi} } \phi(x^\mu)
$, 
with $f(0)=0$.
The identification of  states that differ by a null state reduces the  state space to 
 the zero modes of $\hat{\Phi}$.
 One can also  see this  by noticing   that 
 to compute the inner product of any two states it is sufficient to 
explore the implications of the Fock inner product for the basis $|x^\mu\r$, 
\beq
\l x^\mu_f| \delta(\hat{\Phi}) |x^\mu_i\r = \int d^3 \vec{p} \, 
e^{i ( \vec{x}_f -\vec{x}_i)\cdot \vec{p}- i(t_f-t_i)\hat{R}}= \l \vec{x}_f| e^{-i\hat{R}(t_f -t_i)} |\vec{x}_i\r,
\eeq
which is  the non-relativistic propagator.  What are
 the observables here? Observables are defined by asking that they  commute with $\hat{M}$,
so as not to make physical states unphysical. For example, the operators
$\hat{X}=\hat{x} + i \hat{\lambda} \partial \hat{\Phi}/\partial \hat{p}_x$ and 
$\hat{P}_x=\hat{p}_x - i \hat{\lambda} \partial \hat{\Phi}/\partial \hat{x}$ are 
observables, and one can check that their action reduces to the usual ones in the reduced state space. 
For example,  notice that %writing the physical states as 
%\beq
%|\Psi_{ph}\r = \exp(-\hat{\lambda}\hat{\Phi}) |\varphi\r,
%\eeq
% we can
%see that
\bea
\hat{X} |\Psi_{ph}\r  &=& \exp(-\hat{\lambda}\hat{\Phi}) \hat{x}\, |\varphi\r \nonumber \\
 \hat{P}_x |\Psi_{ph}\r &=& \exp(-\hat{\lambda}\hat{\Phi}) \hat{p}_x \, |\varphi\r
\eea

In the free  parametrized relativistic particle case, the  Fock cohomology has two sectors. 
The ``reduced'' physical space is described by
\bea
\psi_+(p_\mu) &=& \varphi_+(p_i)\,  \left(p_t+\sqrt{ p_i^2 + m^2}\right)e^{- \lambda \Phi }, \\
 \psi_- (p_\mu) &=& \varphi_-(p_i)\,  \left(p_t-\sqrt{ p_i^2 + m^2}\right)e^{- \lambda \Phi }.
\eea
One can check that these two sets of states are physical, not null, and that their difference
is not null. The inner product in the reduced physical space is positive definite: both ``positive''
and ``negative'' energy states have positive norms.

An important   corollary is that 
{the Hadamard Green function, or on-shell amplitude,
\beq
\Delta_1(x -y)= {1 \over (2\pi)^4}\int d^4 k \; \delta(\Phi)\, 
{ e^{-ik(x-y)}} ,
  \eeq
 can be written in the   explicit BRST-Fock
form
\beq
i\Delta_1 (x-y) = \l x^\mu_f| \delta(\hat{\Phi}) |x^\mu_i\r=\l M \! =\!0,c \! =\!0, x^\mu_f |  
M \! =\!0,c \! =\!0, x^\mu_i\r ,
\eeq
with  $ M \equiv \Phi+i p_\lambda$. As was already 
show in reference \cite{Marnelius1993.2}, this amplitude can  also
be rewritten in the  explicit BRST form  
\beq
i\Delta_1 (x-y) =
\l    \psi_{\varphi\! =\! 0 \! =\!  \pi \! =\!   
\eta_0 }| 
\; e^{ [\hat{\cal O}_{NC},\hat{\Omega} ]}\;
         |  \psi'_{\varphi\! =\! 0 \! =\!  \pi \! =\!  \eta_0   }  \r ,
\eeq
where these states are in the Fock 
representation  and  the gauge-fixing is non-canonical. As explained above, it is 
simple to rewrite  this expression  as the BFV path integral in phase space and, after
integration of momenta,  derive the geometric path integral in coordinate space \cite{Gomis}.  This
tells us  that the full-range lapse prescription in the path integral formalism is well grounded
in  a BRST-Fock Hilbert space quantization. 

In the interacting case the analysis of the Fock cohomology is
more complicated.  Null states are still given by the zero modes of $\hat{\Phi}$, which are not
simple to analyze.  In the case of an  electric field with zero divergence, the situation remains simple, since
$\hat{\Phi}$ decouples.  
%The conclusion must be that when this condition is not met, an infinite 
%number of sectors  appears, each containing a different charge.

%%%%%%%%%%%%%%%%%%%%%%%%%%%%%%%%%%%%%%%%%%%%%%%%%%%
\section{Conclusion} Let us summarize our results.
Using reduced phase space quantization leads to 
the loss of  symmetries. Reduced phase 
space quantization is not general enough to be applied to systems with complex phase spaces.

Dirac quantization leads to the Klein-Gordon inner product, but it could
also lead to other possibilities, depending on our choice of hermitean ordering in
the definition of  inner product.
This inner product is not 
positive definite (unless an absolute value is used in the definition of
inner product),  and has to be redefined, since the physical state space is not
really in the original state space.  Time disappears from the formalism,
but can be recovered with the  careful interpretation of the observables.

BRST quantization yields two sectors, one akin to the Dirac state space, with  the 
Klein-Gordon inner product, the other leading to on-shell
amplitudes. This formalism
is still not totally well-defined and the process seems a bit 
{\it ad hoc}.  What happens with the other sectors?  Some questions remain.
It seems, however,  that BRST encodes the Dirac and anti-Dirac approaches to quantization 
(see below).

 In BRST-Fock, gauge-fixing is not need, nor the
redefinition of inner product: the physical states reside already in 
the extended state space---they have finite norms. The arbitrariness in this
approach resides in the definition of the vacuum (i.e., the definition of
annihilation and creation operators).
Two sectors appear in the theory, at least in the free case. The inner product
is positive definite inner product, aside form the null states.  The lapse has full range.
The amplitudes are ``on-shell'', although the states do not satisfy the Dirac condition.
Time remains in the picture.  For instance, in the parametrized non-relativistic
case this quantization approach reduces to the usual non-relativistic quantum mechanics.
This formalism is the  ground and backbone for the path integrals
usually discussed in the literature.  This representation is identical to the ``anti''-Dirac 
one in the zero-ghost sectors in the BRST section.

The different approaches to quantization then essentially lead to two types of amplitudes: symbolically,
\beq
\l \Psi_{\Upsilon=0}^a| \delta(\hat{\Phi})\,  |\Psi_{\Upsilon=0}^b\r , \:\:\:\:\:
 \l \Psi_{\Phi=0}^a| |\overbrace{\delta(\hat{\Upsilon}) \, \{\hat{\Phi}, \hat{\Upsilon}\}}|\,  |\Psi_{\Phi=0}^b\r.
\eeq  
Both of these yield positive norms for the states (although the needed absolute value
does not arise from BRST  in a  natural way). When the constraint can be made into a momentum,
$P\approx 0$, by a canonical transformation, these
amplitudes yield equivalent theories: 
they essentially reduce to 
\beq
\l \Psi_{Q=0}^a| \delta(\hat{P})\,  |\Psi_{Q=0}^b\r , \:\:\:\:\:
 \l \Psi_{P=0}^a| \delta(\hat{Q})  |\Psi_{P=0}^b\r
\eeq  
and they both eliminate the gauge degrees of freedom from the theory.  
Thus, if we had started 
with the constraint $Q\approx0$, we would have ended up with the same
result. If we follow the same logic through with the toy model in which
the constrain is $\Phi=P^2-P_o^2\approx 0$ ($P_o> 0$), we will also find  that the
amplitudes in the two approaches,
\bea
\l \Psi_{Q=Q_o}^a| \delta(\hat{P}^2 &-& P_0^2)\,
|\Psi_{Q=Q_o}^b\r \\
& =& \l
\Psi_{Q=Q_o}^a| \left( \delta(\hat{P}-P_o)/P_o +
\delta(\hat{P}+P_o)/P_o\, \right) |\Psi_{Q=Q_0}^b\r \nonumber ,
\eea
\bea
\l \Psi_{\Phi=0}^a| |\delta(\hat{Q} &-& Q_o)\hat{P} + \hat{P}
\delta(\hat{Q}-Q_o)| \, |\Psi_{\Phi=0}^b\r \\ 
&=& \l \Psi_{P=P_o}^a| \delta(\hat{Q}-Q_o) P_o
|\Psi_{P=P_o}^b\r + \l \Psi_{P=-P_o}^a| \delta(\hat{Q}-Q_o) P_o
|\Psi_{P=-P_o}^b\r \nonumber, 
\eea
yield, at the end,  identical physical spaces---with two separate sectors.
 Notice that, for example,
\bea
\l P  \!  = \! P_o| \delta(\hat{Q}-Q_o) \, | P \! = \! P_o\r &=&  \l P \! = \! P_o| {Q} \! = \! Q_o  \r
\l {Q} \! = \! Q_o | P=P_o\r\nonumber\\
&=& \l {Q} \! = \! Q_o |\delta(\hat{P}-P_o)\, |  {Q} \! = \!  Q_o \r
\eea
so the equivalence is clear.
 In BRST theory, as we just mentioned, 
the two approaches are available, appearing in the different zero-ghost
sectors.

%\ \\
%
%Questions: why are these versions of the relativistic particle not
%usually discussed? What is wrong with the use of the absolute value, 
%in terms of the representations of the Lorentz group?
%What happens in the interacting case, with particle creation? What are the sectors in Fock?
%What happens to the equivalence of the approaches?
%%%%%%%%%%%%%%%%%%%%%%%%
\section*{Acknowledgments}
The author wishes to thank Emil Mottola for lots of  help and useful comments in the course of this research, and  Los Alamos National Laboratory and  UC Davis for
support,  and the Instituto de Matem\'aticas y F\'\i sica Fundamental (IMFF-CSIC) 
for support and valuable discussions.
%, and
%Julianne Chisholm for careful proofreading.

\end{document}